\documentclass[%
prl,twocolumn,
superscriptaddress,
amsmath,amssymb,
aps,
floatfix,
]{revtex4-2}

\usepackage{graphicx}
\usepackage{dcolumn}
\usepackage{bm}
\usepackage{multirow}
\usepackage{braket}
\usepackage{amsfonts}
\usepackage{amssymb}
\usepackage{amsthm}
\usepackage{array}
\usepackage{amsmath}
\usepackage{verbatim}
\usepackage{hyperref}
\usepackage{color}
\usepackage{bbold}
\usepackage{epstopdf}
\usepackage{mathtools}
\usepackage{MnSymbol}
\usepackage{tikz}
\usepackage[export]{adjustbox}
\usepackage{subfigure}
\usepackage{threeparttable}

\newcommand{\affa}{Beijing Academy of Quantum Information Sciences, Beijing 100193, China}
\newcommand{\affb}{Key Laboratory of Atomic and Subatomic Structure and Quantum Control (Ministry of Education), School of Physics, South China Normal University, Guangzhou 510006, China}
\newcommand{\affc}{State Key Laboratory of Low Dimensional Quantum Physics and Department of Physics, Tsinghua University, Beijing 100084, China}
\newcommand{\affd}{Hefei National Laboratory, Hefei 230088, China}

\def\ket#1{\left|#1 \right>}

\begin{document}
	

\title{Mapping Topology-Localization Phase Diagram with Quasiperiodic Disorder Using a Programmable Superconducting Simulator}
\author{Xuegang Li}
\thanks{These authors contributed equally to the work.}
\address{\affa}
\author{Huikai Xu}
\thanks{These authors contributed equally to the work.}
\address{\affa}
\author{Junhua Wang}
\thanks{These authors contributed equally to the work.}
\address{\affa}
\author{Ling-Zhi Tang}
\address{\affb}
\author{Dan-Wei Zhang}
\email{danweizhang@m.scnu.edu.cn}
\address{\affb}
\author{Chuhong Yang}
\address{\affa}
\author{Tang Su}
\address{\affa}
\author{Chenlu Wang}
\address{\affa}
\author{Zhenyu Mi}
\address{\affa}
\author{Weijie Sun}
\address{\affa}
\author{Xuehui Liang}
\address{\affa}
\author{Mo Chen}
\address{\affa}
\author{Chengyao Li}
\address{\affa}
\author{Yingshan Zhang}
\address{\affa}
\author{Kehuan Linghu}
\address{\affa}
\author{Jiaxiu Han}
\address{\affa}
\author{Weiyang Liu}
\address{\affa}
\address{\affd}
\author{Yulong Feng}
\address{\affa}
\author{Pei Liu}
\address{\affc}
\author{Guangming Xue}
\address{\affa}
\address{\affd}
\author{Jingning Zhang}
\email{zhangjn@baqis.ac.cn}
\address{\affa}
\author{Yirong Jin}
\email{jinyr@baqis.ac.cn}
\address{\affa}
\author{Shi-Liang Zhu}
\address{\affb}
\address{\affd}
\author{Haifeng Yu}
\address{\affa}
\address{\affd}
\author{S. P. Zhao}
\address{\affa}
\address{\affd}
\author{Qi-Kun Xue}
\address{\affa}
\address{\affd}
\address{\affc}


\begin{abstract}
We explore topology-localization phase diagram by simulating one-dimensional Su-Schrieffer-Heeger (SSH) model with quasiperiodic disorder using a programmable superconducting simulator. We experimentally map out and identify various trivial and topological phases with extended, critical, and localized bulk states. We find that with increasing disorder strength, some extended states can be first replaced by localized states and then by critical states before the system finally becomes fully localized. The critical states exhibit typical features such as multifractality and self-similarity, which lead to surprisingly rich phases with different types of mobility edges and scaling behaviors on the phase boundaries. Our results shed new light on the investigation of the topological and localization phenomena in condensed-matter physics.
\end{abstract}

\maketitle

{\em Introduction.}---The interplay between topology and disorder has stimulated considerable theoretical~\cite{li2009topological, groth2009theory, guo2010topological, mondragon2014topological, altland2014quantum, titum2015disorder, zhang2020nonhermitian, tang2022topological} and experimental~\cite{meier2018observation, stutzer2018photonic, liu2020topological, lin2022observation, xiao2021observation} interests recently. While strong disorder can drive a trivial material into Anderson localized state~\cite{anderson1958absence}, moderate disorder in many cases leads to a nontrivial state with nonzero topological number and zero-energy edge modes, which is referred to as topological Anderson insulators (TAIs)~\cite{li2009topological, groth2009theory}. Also, depending on the specific types of disorder, the resulting bulk properties may differ considerably. With random disorder in low dimensions, the TAIs are gapless and contain only localized bulk states without mobility edge~\cite{mondragon2014topological, altland2014quantum, meier2018observation}. With quasiperiodic disorder that is deterministic, the critically localized states appear, with unique properties like critical wave function, multifractality, self-similarity, and slow dynamics~\cite{han1994critical, chang1997multifractal, ketzmerick1998covering, takada2004statistics, gong2008fidelity, liu2015localization, zhang2018universal, szabo2020mixed, wang2021many, tang2021localization, xiao2021observation, liu2022anomalous, li2023observation, zhou2023exact, goncalves2023critical, gonifmmode2024short, ceccatto1989quasiperiodic, chandran2017localization, crowley2018quasiperiodic, agrawal2020universality, ohgane2023quasiparticle, natphys2024}, distinct from those of the extended or fully (Anderson) localized states. 

The critical states in quasiperiodic systems have attracted much attention and are often investigated in the framework of the generalized Aubry-Andr\'{e}-Harper (gAAH) model~\cite{han1994critical, chang1997multifractal, ketzmerick1998covering, takada2004statistics, gong2008fidelity, liu2015localization, zhang2018universal, szabo2020mixed, wang2021many, tang2021localization, xiao2021observation, liu2022anomalous, li2023observation, zhou2023exact, goncalves2023critical, gonifmmode2024short}, with a few works concerning the topological phenomena~\cite{liu2015localization, tang2021localization, xiao2021observation}. They have also been a hot topic for the quantum Ising systems~\cite{ceccatto1989quasiperiodic, chandran2017localization, crowley2018quasiperiodic, agrawal2020universality, ohgane2023quasiparticle}. On the other hand, there are many studies of the topological states~\cite{tang2022topological, cai2019observation, longhi2020topological, roy2021reentrant, lu2022exact, lu2023robust, katz2024observing, splitthoff2024gatetunable} based on the Su-Schrieffer-Heeger (SSH) model~\cite{su1979solitons, zhang2018topological, cooper2019topological}. In particular, topological phase diagrams with quasiperiodic disorder have been theoretically studied recently~\cite{tang2022topological, lu2022exact, lu2023robust}, but it is not clear whether the critical states may exist and fundamentally change the properties of the phase diagram.

Quantum simulators provide a valuable tool for the studies of various problems ranging from condensed-matter physics and quantum chemistry, to high-energy physics and cosmology~\cite{georgescu2014quantum, daley2022practical}, even in the present NISQ era~\cite{preskill2018quantum}. Many problems are often hard to study in real materials. For instance, topological~\cite{cai2019observation, shi2023quantum} and localization~\cite{karamlou2022quantum} states and quantum scars~\cite{zhang2023many} have been simulated using superconducting quantum processors. The TAIs have been studied and demonstrated in the ultracold atom~\cite{meier2018observation} and photonic~\cite{stutzer2018photonic, liu2020topological} systems by tracing along particular parameter paths across the phase boundaries. Nevertheless, the experimental study and mapping of an overall topological phase diagram remain largely unexplored, especially when associated with different localization properties of the bulk states. 

\begin{figure}[t]
\includegraphics[width=0.4\textwidth]{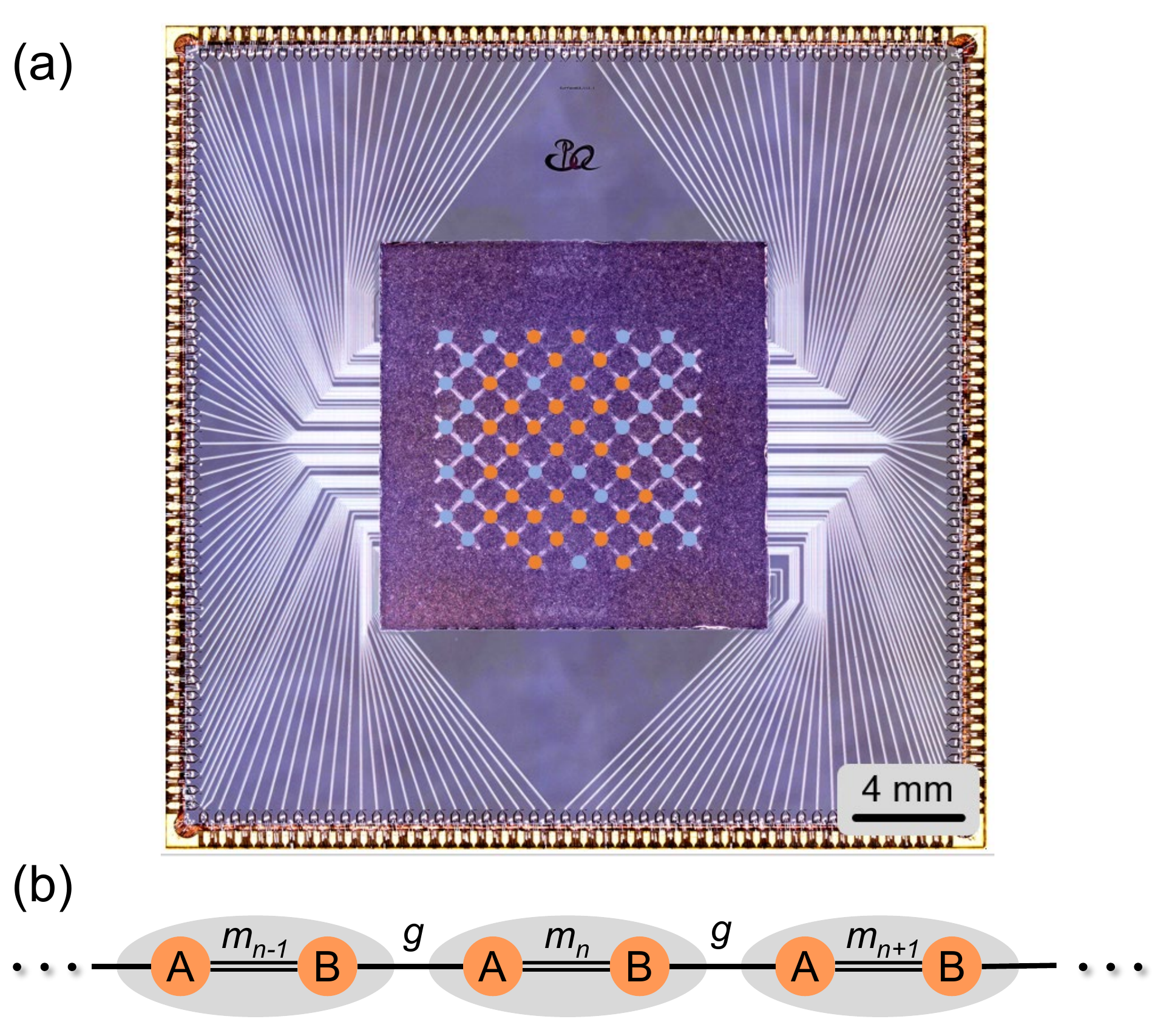}
\caption{(a) Photograph of the programmable superconducting simulator. The circuits on the top half-transparent sapphire substrate are on the back side seen as a grid image and color circles are added at the nodes to indicate qubits, with those in orange forming a 32-qubit chain used in this work. (b) Schematic of the gSSH model. The inter-cell tunneling (single lines) is homogeneous, while the intra-cell tunneling (double lines) is modulated with quasiperiodic disorder.}
\label{fig:Figure_1}
\end{figure}

In this work, we present an experiment of mapping the topology-localization phase diagram with quasiperiodic disorder using a programmable superconducting quantum simulator~\cite{supplementary} developed by flip-chip technology~\cite{li2021vacuum}. The simulator has 63 tunable qubits and 105 tunable couplers arranged in 2D square lattice with extremely small next-nearest-neighbor (diagonal) coupling on the order of 2$\pi \times$30 kHz, allowing engineering of Hamiltonians conveniently and accurately by controls over qubits and couplers via programming. A chain of 32 qubits is used in this work as shown in Fig.~\ref{fig:Figure_1}(a). By simulating the one-dimensional SSH model with quasiperiodic disorder, we experimentally map out a complete topology-localization phase diagram and identify various trivial and topological phases having extended, critical, and localized bulk states. With increasing disorder strength, some extended states can be first replaced by localized states and then by critical states, and finally, the system becomes fully localized. We find that the critical states have typical features such as multifractality and self-similarity. Their presence changes the structure of the phase diagram as compared to the previous works~\cite{tang2022topological, lu2022exact, lu2023robust}, leads to new types of mobility edges, and affects the scaling behaviors and critical exponents on the phase boundaries.

{\em Model.}---The SSH model is originally proposed to describe spin-less fermions in dimerized chains [see Fig.~\ref{fig:Figure_1}(b)]~\cite{su1979solitons, zhang2018topological, cooper2019topological}. It is the simplest model widely used in the studies of topological states~\cite{cai2019observation} and TAI states with random~\cite{meier2018observation} and quasiperiodic~\cite{tang2022topological, longhi2020topological, roy2021reentrant, lu2022exact} disorders. For a dimerized chain with $L$ lattice sites or $N_c$ = $L/2$ unit cells, the model with quasiperiodic disorder reads
\begin{eqnarray}
\hat H_{\rm gSSH}/\hbar=\sum_{n=1}^{N_c}m_n\hat a_{n}^\dag\hat b_{n}+g\sum_{n=1}^{N_c-1}\hat a_{n+1}^\dag\hat b_{n}+{\rm h.c.},
\label{ssh1}
\end{eqnarray}
where $\hat a_n$ ($\hat b_n$) is the fermionic annihilation operator for the $A$ ($B$) site in the $n$-th unit cell. This generalized SSH (gSSH) model can be mapped by the Jordan-Wigner transformation to a spin-$\frac{1}{2}$ system that can be conveniently simulated by a superconducting qubit chain~\cite{cai2019observation, yan2019strongly}. We consider a simple case where the inter-cell tunneling $g$ is homogeneous and quasiperiodic disorder is introduced in the intra-cell tunneling term:
\begin{eqnarray}
m_n=m+W\cos\left(2n\pi\beta+\delta\right).
\label{ssh2}
\end{eqnarray}
Here $W$ quantifies the disorder strength, $\beta$ = $(\sqrt{5}-1)/2$ is an irrational number~\cite{ketzmerick1998covering}, and $\delta$ is an arbitrary phase. $\delta$ can be used to generate different disorder realizations for finite-size systems, which are averaged to recover the thermodynamic results. The disorder in this simple form is favorable for the observations of TAI and critical states and the determination of phase boundaries~\cite{supplementary}. The topological and localization properties of the 1D disordered chiral system can be characterized by the real-space winding number $N_{\rm w}$ and the spectrally averaged inverse participation ratio $\overline{\rm IPR}$, namely $N_{\rm w}$ = 0 and 1 for topological trivial and nontrivial states, and $\overline{\rm IPR}$ = 0 and $>$ 0 for extended and localized states, respectively~\cite{tang2022topological}.

{\em Experiment.}---We use 32 qubits from the simulator forming a 1D system for the simulation of the gSSH model, see Fig.~\ref{fig:Figure_1}. The active qubits are biased to a common frequency $\omega_{\rm ref}=2\pi\times5.495$ GHz. The nearest-neighbor couplings are tuned to yield an inter-cell tunneling strength of $g=2\pi\times1.5$ MHz and the ratio $m/g$ ($W/g$) varied in the range $\left[0,2\right]$ ($\left[0,4\right]$). In the experiment, we first align the frequencies of 32 active qubits, then parameterize the coupling strength of each nearest-neighbor qubit pair, and finally compensate for the coupler-induced dispersive shifts. By a fast control procedure, we are able to run the experiment with a switching rate of 18 target Hamiltonians per hour. The statistical fidelity, i.e. $F(p,q)=\sum_x\sqrt{p_xq_x}$ between the measured and ideal population distribution~\cite{yan2019strongly} in the dynamical evolution averaged over different target Hamiltonians is over 91\% within the maximum evolution time. 

We quantitatively extract the topological and localization properties, i.e. $N_{\rm w}$ and $\overline{\rm IPR}$, from the time-averaged expectation values of chiral displacement operator and survival probability in quantum-walk experiments:
\begin{eqnarray}
C_t&=&\frac{2}{t}\int_0^t\left\langle\psi\left(t'\right)\left|\hat\Gamma\hat X\right|\psi\left(t'\right)\right\rangle dt', \\
S_t&=&\frac{1}{t}\int_0^t\left|\left\langle l_0|\psi\left(t'\right)\right\rangle\right|^2dt'.
\end{eqnarray}
Here $\hat\Gamma\equiv\sum_n\left(\hat a_n^\dag\hat a_n-\hat b_n^\dag\hat b_n\right)$ and $\hat X\equiv\sum_nn\left(\hat a_n^\dag\hat a_n+\hat b_n^\dag\hat b_n\right)$. The quantum-walk dynamics is governed by $\left|\psi\left(t\right)\right\rangle=\exp\left(-i \hat H_{\rm gSSH}t/\hbar\right)\left|l_0\right\rangle$ where $\ket{l_0}$ is the initial Fock state with only the $l_0$-th site excited. $N_{\rm w}$ can be obtained from $C_t$~\cite{cardano2017detection} while $\overline{\rm IPR}=\lim_{t\rightarrow\infty}\overline{S}_t$ with $\overline{S}_t=\frac{1}{L}\sum_{l_0}S_t(l_0)$~\cite{supplementary}. In the experiment, we construct 8 disorder realizations and 4 initial single-excitation states for a given target Hamiltonian. The dynamical properties of $C_t$ and $S_t$ for the 8$\times$4 quantum-walk instances with given parameters $W/g$ and $m/g$ are measured to obtain the instance-averaged $\overline{C}_t$ and $\overline{S}_t$. Although the evolutions of $C_t$ and $S_t$ depend on the specific disorder realization and initial state, $\overline{C}_t$ and $\overline{S}_t$, after taking average over the 32 instances, converge to $N_{\rm w}$ and $\overline{\rm IPR}$, respectively.

\begin{figure}[t]
	\includegraphics[width=0.48\textwidth]{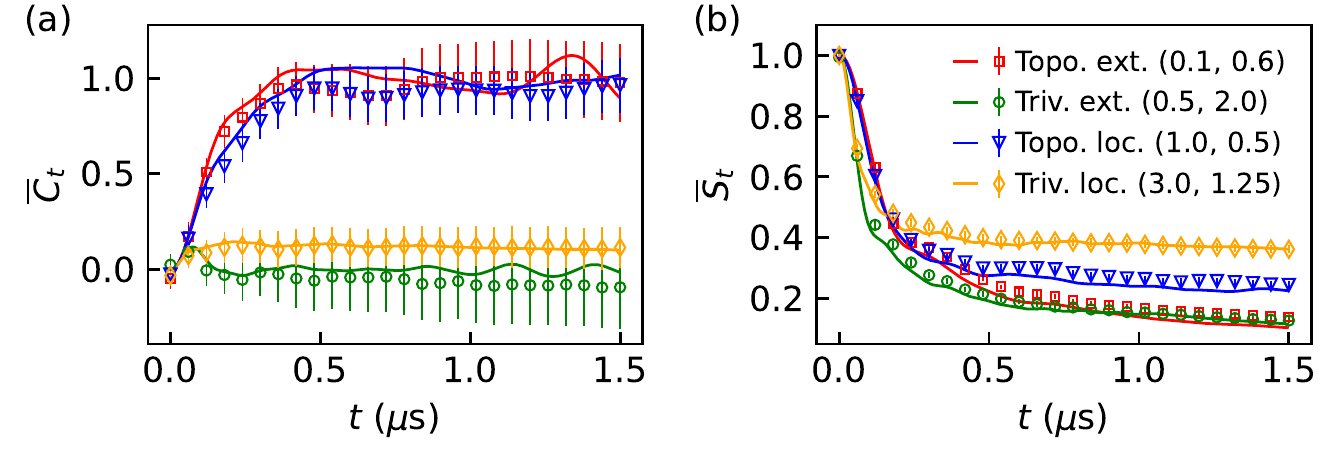}
	\caption{Dynamical properties of (a) averaged chiral displacement and (b) survival probability in different topological and localization phases with parameters $(W/g, m/g)$ indicated in the legend. Here symbols and lines are the experimental and numerical results, respectively. For the experimental results, the error bars represent the standard error of the mean arising from the sampling error in the projective measurement with 1024 repetitions.}
	\label{fig:Figure_2}
\end{figure}

Figure~\ref{fig:Figure_2} shows the measured $\overline{C}_t$ and $\overline{S}_t$ for four ($W/g$, $m/g$) parameter sets, corresponding to different topological and localization properties in $W$-$m$ plane described below. Quantum walks with these parameters show that the edge excitations in topological phase remain largely at the edge in time evolution no matter the bulk states are localized or extended, while in trivial case the excitations will remain at the edge for localized state and spread to all qubits for extended state~\cite{supplementary}. In Fig.~\ref{fig:Figure_2}(a), we find that the quantity  $\overline{C}_t$ almost converges to the corresponding values in the long-time limit, say $N_{\rm w}$ = 1 (0) for topological (trivial) phases, while it is less straightforward for the results concerning localization property as shown in Fig.~\ref{fig:Figure_2}(b). Ideally, $\overline{S}_t$ should vanish for systems of infinite size at infinite time in the extended state. In the present case, it remains finite due to the finite size and time in experiment. However, we observe that $\overline{S}_t$ decreases with time at the rates clearly higher for extended states than those for localized states. The values of $\overline{S}_t$ near the end of evolution still appear separate enough to discriminate the localized states from the extended ones, which can be verified by numerical simulations in the corresponding cases with infinite size and time. 

\begin{figure}[t]
	\includegraphics[width=0.4\textwidth]{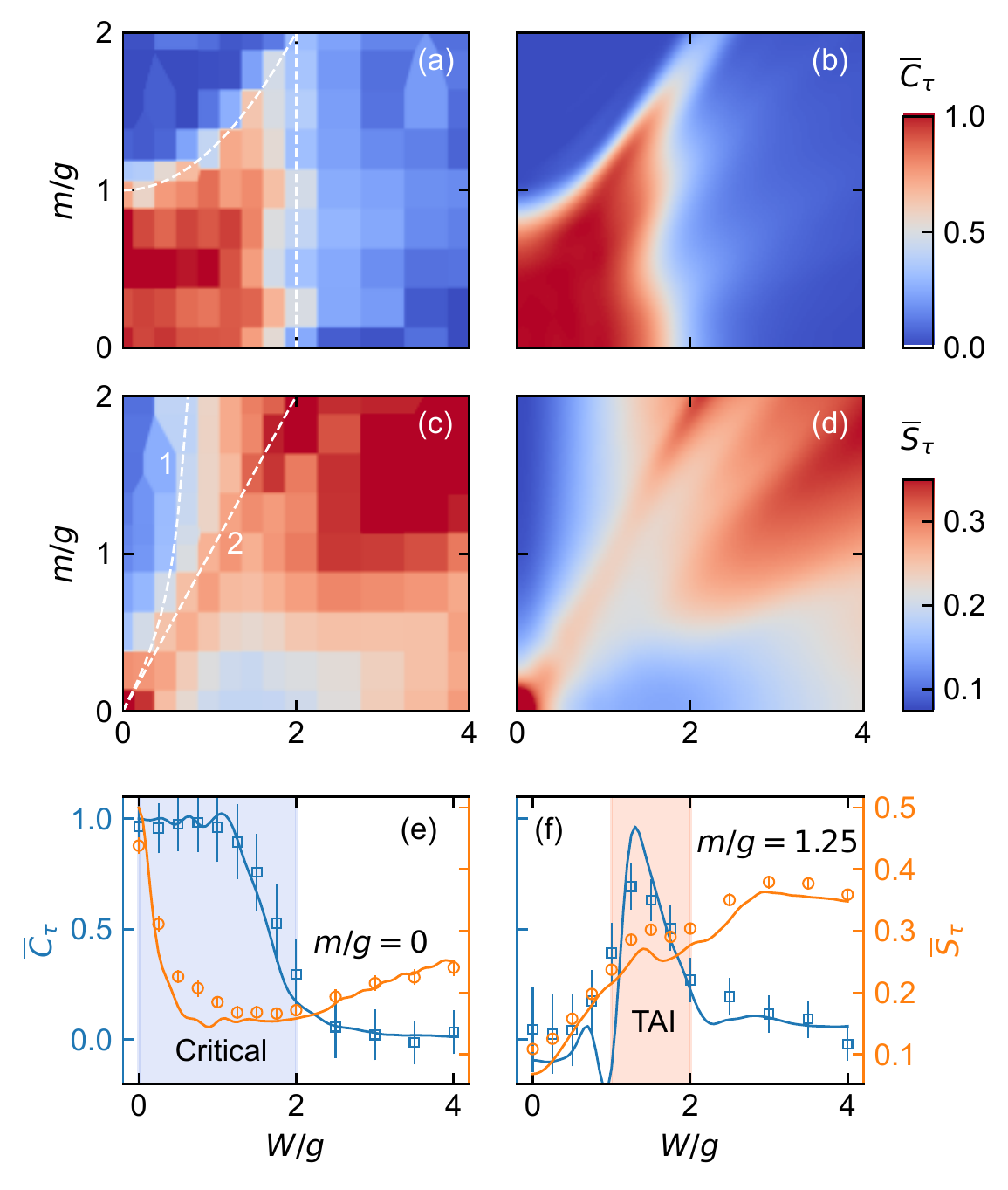}
	\caption{Topology-localization phase diagram seen from (a) experimental and (b) numerical averaged chiral displacement $\overline{C}_\tau$; (c) experimental and (d) numerical averaged survival probability $\overline{S}_\tau$. Each plaquette in (a) and (c) represents one data point on the parameter plane. White dashed lines indicate calculated phase boundaries in Fig.~\ref{fig:Figure_4}(d) considering extended qubit number. Two representative cross-sections of the parameter space are shown in (e) and (f) with symbols and lines being the experimental and numerical results, and shaded areas indicating pure critical states and TAI regimes. The colorbar in (c) and (d) is set for better contrast with the data near origin out of range as can be seen in (e).}
	\label{fig:Figure_3}
\end{figure}

{\em Phase diagram.}---To map out the phase diagram, we sweep the model parameters in $W$-$m$ plane and the experimental results are presented in Figs.~\ref{fig:Figure_3}(a) and (c). Here we use the measured values at the maximum evolution time $\tau=1.5~\mu$s, denoted by $\overline{C}_\tau$ and $\overline{S}_\tau$, as the experimental quantities for evaluating topological and localization properties. We choose 114 points unevenly distributed in the $W$-$m$ plane, and obtain the real-space winding number $N_{\rm w}$ ($\overline C_\tau$) and the spectrally averaged inverse participation ratio $\overline{\rm IPR}$ ($\overline S_\tau$) in the long-time limit. Numerical results calculated~\cite{johansson2012qutip} using $L$ = 32 and open boundary condition without free fitting parameters are shown in Figs.~\ref{fig:Figure_3}(b) and(d), demonstrating good agreements between experiment and theory.

The experimental results in Figs.~\ref{fig:Figure_3}(a) and (c) reveal remarkably rich phases for the gSSH model. First of all, there exist two distinct regions in Fig.~\ref{fig:Figure_3}(a), namely topologically trivial and nontrivial, as separated by dashed lines for $N_{\rm w}$ ($\overline{C}_\tau$) values close to 0 and 1, respectively. Secondly, as can be seen from increasing $\overline{\rm IPR}$ ($\overline{S}_\tau$) in Fig.~\ref{fig:Figure_3}(c), the bulk states change monotonically from extended to localized ones for $m/g$ $\sim$ 2 as the disorder parameter $W$ increases. However, for $m/g$ around 1, we find that $\overline{\rm IPR}$ ($\overline{S}_\tau$) first increases for $W/g$ up across the dashed straight line (phase boundary 2) and then decreases before finally increasing to more localized state. We also observe a clear opposite trend near $m/g$ = 0. Namely, the bulk states change from more localized to less localized ones with increasing $W$, and further increasing $W$ leads to more localized state again (note that at $m/g = W/g$ = 0 the system is completely localized due to the disconnectivity of the chain). Figures~\ref{fig:Figure_3}(e) and (f) show $N_{\rm w}$ ($\overline{C}_\tau$) and $\overline{\rm IPR}$ ($\overline{S}_\tau$) versus $W$ for $m/g$ = 0 and 1.25, respectively, in which the pure critical states and TAI regimes are indicated.

\begin{figure}[t]
	\includegraphics[width=0.48\textwidth]{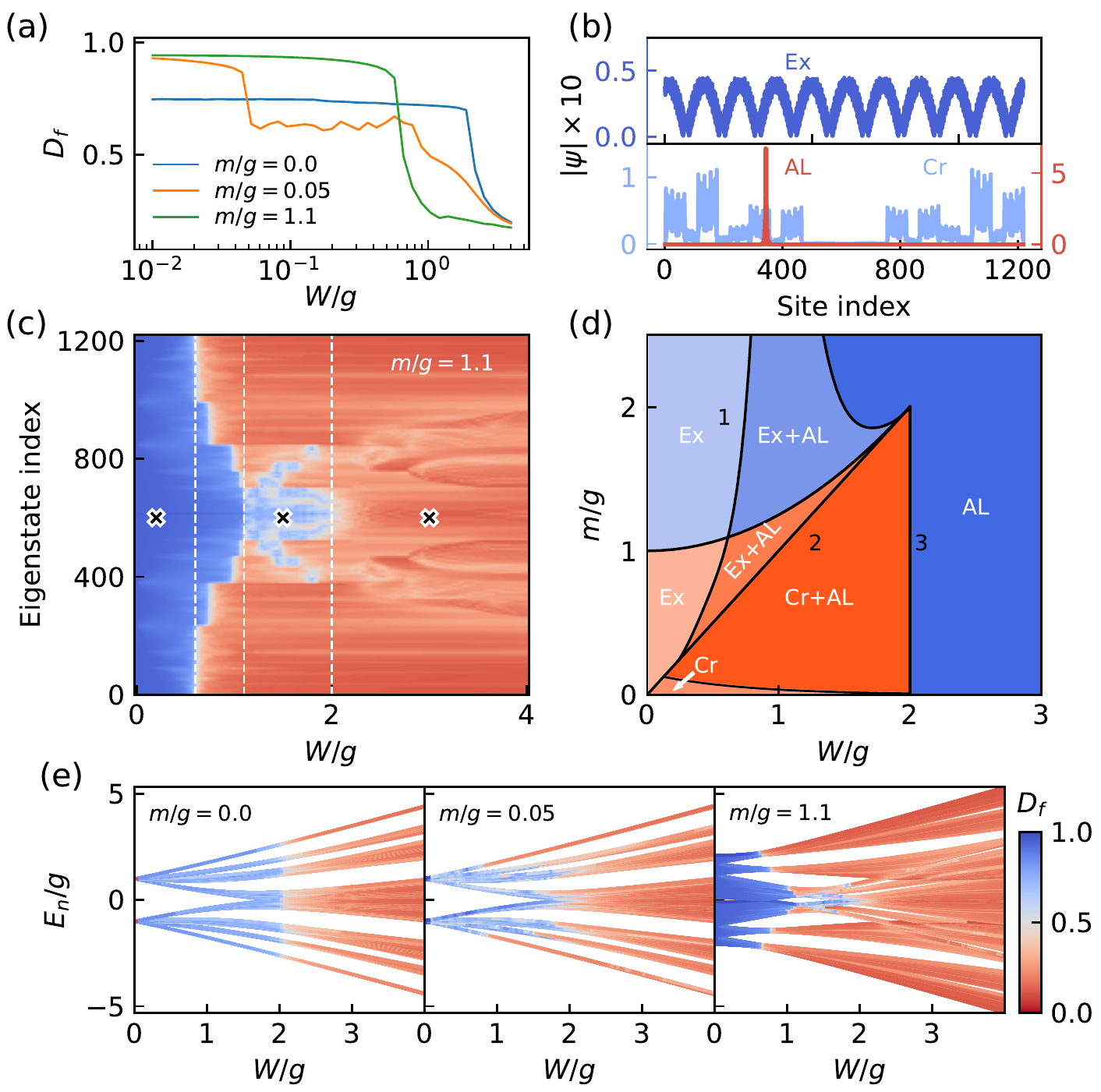}
	\caption{(a) Fractal dimension $D_f$ versus $W/g$. (b) Wave functions for $m/g$ = 1.1 and three different $W/g$ marked by crosses from left to right in (c), corresponding to the extended, critical, and localized states respectively. (c) Eigenstate $D_f$ vs $W/g$ for $m/g$ = 1.1, sharing the same colorbar in (e). Three vertical lines from left to right indicate the points on phase boundaries 1, 2, and 3 of bulk-state transitions in (d), respectively. (d) Phase diagram. Orange (blue) color indicates  topologically nontrivial (trivial) phase, with increasing darkness for more localized states. In (b) and (d), Ex, Cr, and AL denote extended, critical, and Anderson localized states, respectively. (e) Eigenstate $D_f$ versus $W/g$ for three $m/g$ values. All results are calculated using Eqs.~(\ref{ssh1}) and (\ref{ssh2}) with $N_c$ = 610 and periodic boundary condition.}
	\label{fig:Figure_4}
\end{figure}

To further understand the experimental phase diagram in Fig.~\ref{fig:Figure_3}, we turn to the numerical results calculated using Eqs.~(\ref{ssh1}) and (\ref{ssh2}) with larger $L$ = 1220 and periodic boundary condition for more general circumstances closer to the thermodynamic limit. Figure~\ref{fig:Figure_4}(a) shows the result of $-\ln\overline{\rm IPR}/\ln L$ as a function of $W/g$, which corresponds to the fractal dimension $D_f$ when $L$ tends to infinity~\cite{RevModPhys.80.1355, wang2021many}. For $m/g$ = 0, the system is seen to be critically localized for $0 < W/g < 2$ in terms of $D_f\sim$ 0.75. Surprisingly, a small increase of $m/g$ from 0 to 0.05 can lead to an abrupt change of bulk state property. The system in this case changes from extended ($D_f\sim$ 1), to critically localized ($D_f\sim$ 0.6), and finally to fully localized ($D_f\sim$ 0) states as $W$ increases. For further increased $m/g$ = 1.1, the extended state persists to larger $W/g$ before finally turns to the fully localized state.

The critical states have unique properties compared to the extended and fully (Anderson) localized states. They have distinct critical wave functions as shown in Fig.~\ref{fig:Figure_4}(b), and are characterized by multifractality as seen from $D_f$ and self-similarity\cite{ketzmerick1998covering, natphys2024} as described in~\cite{supplementary}. Figure~\ref{fig:Figure_4}(c) shows the eigenstate $D_f$ against $W/g$ for $m/g$ = 1.1, with the values of $D_f$ for critical state lying between those for the extended and Anderson localized states. We see that the whole region is occupied by the extended states for small $W/g$. As $W/g$ increases, some extended states are replaced by the Anderson localized states, and further by the critical states resulting in the absence of the extended states, which correspond to the points on phase boundaries 1 and 2 in Fig.~\ref{fig:Figure_3}(c), respectively. Following this there exists an increase of the critical states against Anderson localized states in a narrow range of $W/g$. Finally, the Anderson localized states expand in number and become the only bulk states for $W/g$ above 2. A complete phase diagram is shown in Fig.~\ref{fig:Figure_4}(d)~\cite{supplementary}. It is worth noting that the pure critical phase is discovered in the present experiment partly due to the finite size effect that enlarges the parameter ranges in the phase diagram [compare Fig.~\ref{fig:Figure_3}(c) and Fig.~\ref{fig:Figure_4}(d)].

{\em Discussion.}---The simulated gSSH model has some interesting relations with the gAAH and Ising models, either in limiting cases~\cite{supplementary} or under exact mapping~\cite{note}. For example, with $m \gg g, W$, the gSSH chain can be reduced to two sub-AAH chains, with the average on-site potentials of $m$ and $-m$, respectively, and the intra- and inter-chain coupling of $g/2$. The inter-chain coupling can be neglected due to large $m$. The disorder $W/g$ drives the extended-to-localized transition inside each AAH chain, with the critical value of $W/g$ = 1, which corresponds to phase boundary 1 in Fig.~\ref{fig:Figure_3}(c) and Fig.~\ref{fig:Figure_4}(d) for large $m$. In the case of $g \gg m, W$, relating to the phase diagram near origin, the gSSH chain can be reduced to two off-diagonal AAH chains with the on-site potentials homogeneously shifted to $g$ and $-g$, such that the inter-chain tunneling terms, with the tunneling strength proportional to $m$ and $W$, can be safely neglected. The extended-to-critical transition occurs as $W/m$ increases, with the critical line of $m = W$~\cite{chang1997multifractal}, corresponding to phase boundary 2 near the origin in Fig.~\ref{fig:Figure_3}(c) and Fig.~\ref{fig:Figure_4}(d).

The Ising model with quasiperiodic disorder exhibits some magnetic phase boundaries in form similar to those in Fig.~\ref{fig:Figure_4}(d)~\cite{chandran2017localization}. Following the discussion there, we consider a 2D tight-binding model on a square lattice, with each row described by the SSH model~\cite{supplementary}. The intra-cell quasiperodic disorder is introduced by adding complex couplings over diagonal sites between adjacent rows within the same unit cell. For $m$ = 0, the square lattice decouples to two honeycomb lattices, with a flux of 2$\beta$ penetrating each hexagon and the tunneling strengths of the three bonds connected to the same site of $W/2$, $W/2$ and $g$, respectively. As discussed in Ref.~\cite{chandran2017localization}, the system possesses extra rotational symmetry when $W/g$ = 2, leading to a critical-to-localized phase transition. The gSSH model with $m$ = 0 exhibits energy independent critical-to-localized transition at $W/g$ = 2, as shown in the left panel of Fig.~\ref{fig:Figure_4}(e). However, this simple picture is only valid for $m$ = 0, while increasing $m$ introduces coupling between the two honeycomb lattices, rendering energy-dependent critical-to-localized transition. This yields generalized mobility edges which separate critical and localized states in the regime of moderate disorder, as shown in the middle panel of Fig.~\ref{fig:Figure_4}(e). As $m$ further increases, the regime of critical states shrinks to the vicinity of the zero-energy gap, while mobility edges between extended and localized states emerge, which are shown in the right panel of Fig.~\ref{fig:Figure_4}(e). In this case, the zero-energy gap is almost closed at the disorder-free limit and reopened in the moderate disorder regime, which indicates the existence of the TAI~\cite{tang2022topological}.

The boundaries separating topologically trivial (blue) and nontrivial (orange) phases in Fig.~\ref{fig:Figure_3}(a) and Fig.~\ref{fig:Figure_4}(d) can be obtained analytically by considering the inverse localization length of zero-energy modes with open boundary condition~\cite{supplementary}. They are identified as $m=g+W^2/4g$ and $W=2g$ for $m>W$ and $m<W$, respectively. By analyzing the scaling behaviors of the inverse localization length and energy gap at half filling~\cite{mondragon2014topological, chandran2017localization}, we find their critical exponents $\nu$ and $z$ to be $\nu=z=1$ for $m>W$, and $\nu=1$ but a non-universal $z$ varying around 1.5 for $m<W$. These results show that the topological transitions across the two boundaries above and below the diagonal line $m=W$ belong to two different classes. Notably, in the regions above and below the diagonal line, the critical states are absent and present, which implies that the non-universal critical exponents may result from the transition between the critical and localized states.

Finally, we mention that both the energy spectra and wave functions of the critical states exhibit perfect self-similarity in the gSSH model under periodic and open boundary conditions~\cite{supplementary}. Under open boundary condition, some  eigenstates will appear lying between the subbands of the fractal spectrum and become localized at the edges. These in-gap localized states are not protected by the chiral symmetry, thus different from the topological zero-energy edge states.

{\em Conclusion.}---We have experimentally mapped out a topology-localization phase diagram by simulating the gSSH model and demonstrated the potential of NISQ devices for solving fundamental problems in condensed-matter physics. Topologically trivial and nontrivial phases with (mixed) extended, critical, and localized bulk states have been found, offering a rich phase diagram valuable for further studies. 

We acknowledge supports from the National Natural Science Foundation of China (Grant Nos. 92365206, 12104055, 12104056, and 12174126), Innovation Program for Quantum Science and Technology (No. 2021ZD0301802), and Guangdong Basic and Applied Basic Research Foundation (Grant No. 2024B1515020018).

\bibliography{Topological_Anderson_insulators}

\begin{thebibliography}{58}
\bibitem [{not()}]{note}%
  \BibitemOpen
  \href@noop {} {}\bibinfo {note} {Similar to the discussion in Ref. [31], by
  introducing the Majorana representation and then mapping the Majorana
  fermions to spin operators via the Jordan-Wigner transformation, the gSSH
  model can be mapped exactly to two transverse-field Ising models (TFIMs),
  with the transverse field quasiperiodically modulated.}\BibitemShut {Stop}%
\end{thebibliography}

\end{document}


\title{Supplemental Material for ``Mapping Topology-Localization Phase Diagram with Quasiperiodic Disorder Using a Programmable Superconducting Simulator"}

\author{Xuegang Li}
\thanks{These authors contributed equally to the work.}
\address{\affa}
\author{Huikai Xu}
\thanks{These authors contributed equally to the work.}
\address{\affa}
\author{Junhua Wang}
\thanks{These authors contributed equally to the work.}
\address{\affa}
\author{Ling-Zhi Tang}
\address{\affb}
\author{Dan-Wei Zhang}
\email{danweizhang@m.scnu.edu.cn}
\address{\affb}
\author{Chuhong Yang}
\address{\affa}
\author{Tang Su}
\address{\affa}
\author{Chenlu Wang}
\address{\affa}
\author{Zhenyu Mi}
\address{\affa}
\author{Weijie Sun}
\address{\affa}
\author{Xuehui Liang}
\address{\affa}
\author{Mo Chen}
\address{\affa}
\author{Chengyao Li}
\address{\affa}
\author{Yingshan Zhang}
\address{\affa}
\author{Kehuan Linghu}
\address{\affa}
\author{Jiaxiu Han}
\address{\affa}
\author{Weiyang Liu}
\address{\affa}
\address{\affd}
\author{Yulong Feng}
\address{\affa}
\author{Pei Liu}
\address{\affc}
\author{Guangming Xue}
\address{\affa}
\address{\affd}
\author{Jingning Zhang}
\email{zhangjn@baqis.ac.cn}
\address{\affa}
\author{Yirong Jin}
\email{jinyr@baqis.ac.cn}
\address{\affa}
\author{Shi-Liang Zhu}
\address{\affb}
\address{\affd}
\author{Haifeng Yu}
\address{\affa}
\address{\affd}
\author{S. P. Zhao}
\address{\affa}
\address{\affd}
\author{Qi-Kun Xue}
\address{\affa}
\address{\affd}
\address{\affc}

\maketitle

\tableofcontents

\section{Device and measurement setup}

\subsection{Device design}

\begin{figure}
	\includegraphics[width=0.49\textwidth]{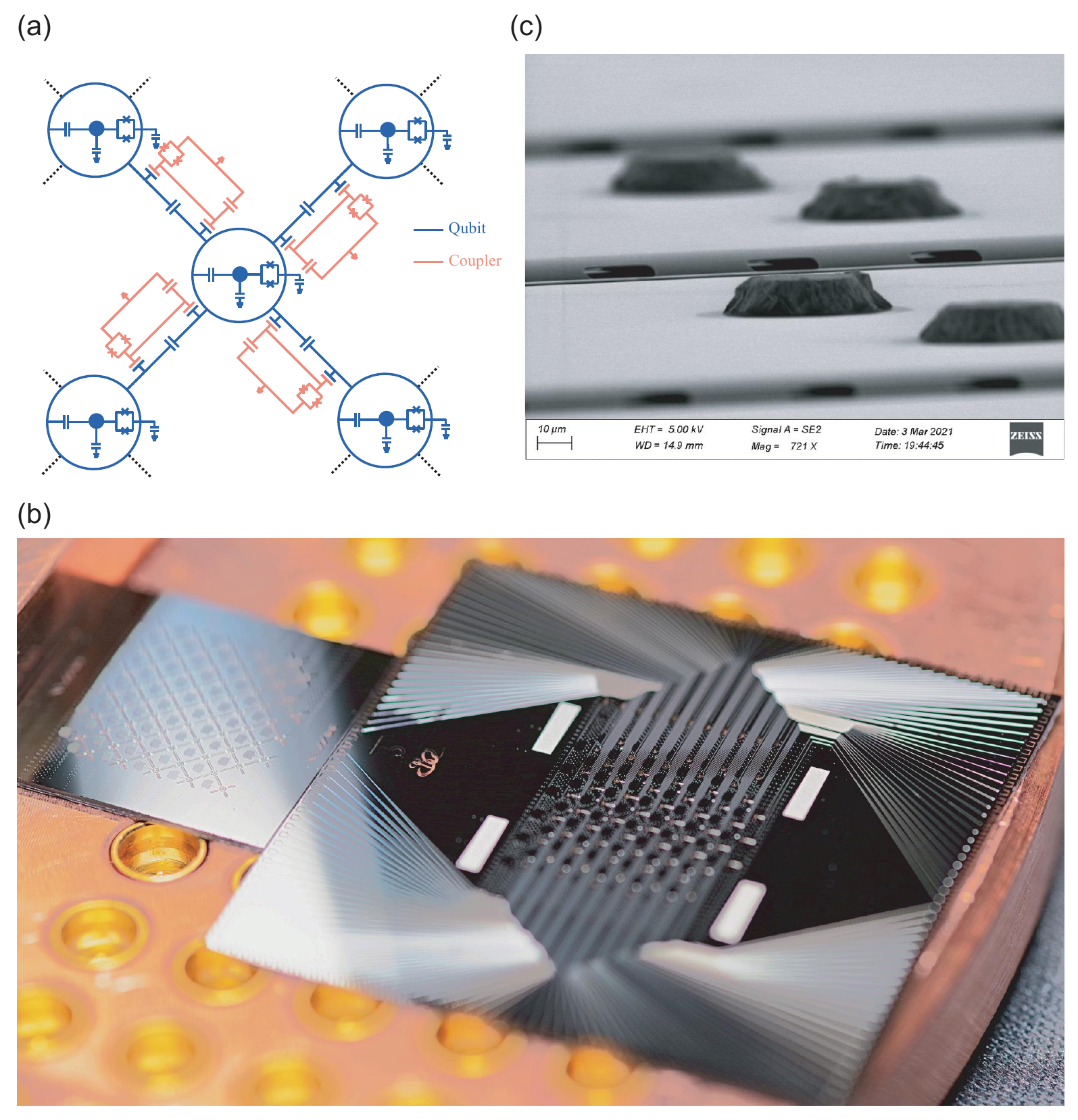}
	\caption{\label{fig:sample} The programmable superconducting quantum processor fabricated using the flip-chip technology. (a) Schematic circuit of a unit cell of the device. Each flipmon-type  qubit (blue) is capacitively coupled with four couplers (orange) and four adjacent qubits. The frequencies of all qubits and couplers can be tuned by the flux of SQUID loops. (b) Photograph of the device with the carrier on the right (3$\times$3 cm$^2$ in size) and top chip on the left (1.5$\times$1.5 cm$^2$ in size). (c) Photograph of the tunnel-like air-bridge and indium bumps.}
\label{fig:device}
\end{figure}

The programmable superconducting quantum processor used in this work consists of 63 tunable qubits and 105 tunable couplers. The qubits are arranged in a 2-dimensional square lattice with a coupler between each of the two nearest-neighbor qubit pairs. A schematic of the circuit of the unit cell is shown in Fig.~\ref{fig:device}(a). The qubits adopt "flipmon" design, which is considered for advantages including small diagonal next-nearest neighbor coupling, improved vacuum energy participation ratio, reduced unwanted crosstalk, more freedom of circuit wiring, and natural compatibility with flip-chip technology~\cite{flipmonpaper}. 

Each qubit includes an asymmetric SQUID, with a control line grounded nearby for tuning its frequency with persistent bias or fast DC pulse. The control line also acts as a microwave drive line utilizing its weak capacitive coupling to the qubit. A meandered quarter-wave coplanar waveguide (CPW) resonator is dispersively coupled to each qubit for readout, and up to six resonators share a common bandpass filter for suppression of the Purcell effect. We put the resonator frequencies far under the qubit's idle frequencies, in the range of 4.1 - 4.6\,GHz. The tunable coupler is a grounded qubit between two neighboring qubits. The tunable couplers have much higher frequencies (about 8\,GHz) at their optimal points. By choosing appropriate qubit and coupler frequencies, the effective coupling strength between adjacent qubits can be continuously tuned from positive to negative~\cite{PhysRevApplied.10.054062} and thus can be turned off.

Figure~\ref{fig:device}(b) shows the photograph of the device. It contains a top chip, where all elements with high-quality factors, including qubits, couplers, and readout resonators, are allocated, and a carrier chip, with other elements, including the bandpass Purcell filters and the control lines. All the Purcell filters and control lines are covered with tunnel-like air-bridges, which are shown in Fig.~\ref{fig:device}(c). Those bridges form Faraday cages that prevent the leakage of electromagnetic fields, thus protecting the qubits and couplers from coupling to spurious fields.

\subsection{Fabrication}

\begin{figure*}
	\centering
	\includegraphics[scale=0.3]{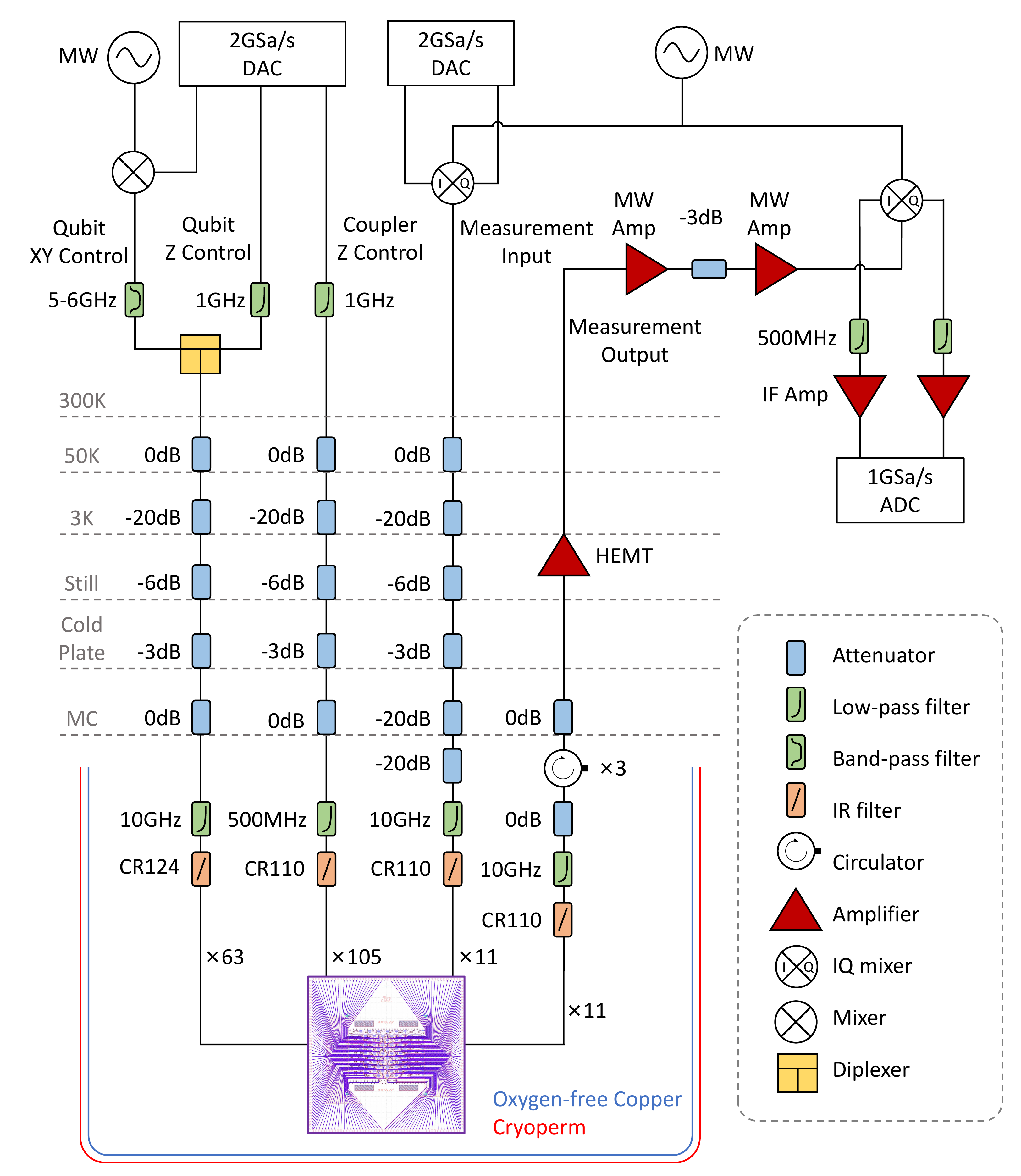}
	\caption{Schematic of the measurement system, including cryogenic and room temperature wirings, various microwave devices, measurement electronics, and electromagnetic shielding.}
	\label{fig:wiring}
\end{figure*}

The carrier and the top chip were fabricated separately with almost the same processes, except that the preparations of Josephson junctions were required for the top chip. About 200\,nm thick tantalum (Ta) films were first deposited on a pre-annealed sapphire wafer by sputtering. The base circuits, including all the CPW transmission lines, resonators, and capacitors, were then defined by laser direct writing lithography (DWL). They were transferred to the Ta film using reactive ion etching (RIE) with $\mathrm{SF_6}$ gas. Before and after the etching process, oxygen ashing was performed in order to remove the residual photoresist. The wafers were then immersed in N-methyl-pyrrolidone (NMP) for several hours, followed by ultrasonic cleaning. The $\mathrm{Al/AlO_{x}/Al}$ junctions were fabricated using the standard Dolan-bridge shadow evaporation technology~\cite{dolan1977offset}. To do so, the junction regions were defined by electron beam lithography with double-layer resists. After development, Dolan-bridges were formed with under-cut structures. Subsequently the wafers were transferred to an ultra-high vacuum four-chamber electron beam evaporation system (AdnanoTek$^\circledR$ JEB-4), wherein about 17\,nm Al was deposited with a tilted angle of 40 - 60 degrees to form the bottom electrodes. They were transferred into the oxidation chamber to form a thin $\mathrm{AlO_x}$ barrier layer and finally back for the deposition of Al top electrodes of about 19\,nm in thickness. Right before the deposition, an in-situ argon ion milling was used to cleanse the surface of the Ta films.

After deposition, the wafers were soaked in NMP bath of 80$^{\circ}$C for at least two hours, with a gentle ultrasonic for thorough lift-off. The tunnel-like bridges were fabricated following the reflowing process~\cite{Chen2014Fabrication} with two differences. First, we increased the thickness of the air-bridge film to 500\,nm to ensure enough structural strength to sustain subsequent processes. Second, the area for wet etching was 6~$\mathrm{\mu m}$ wide surrounding the air-bridges, which ensured that the bridges were separated from the excess aluminum films. Another lift-off was performed similar to that in the Josephson junction process but without ultrasonic. As there was quite an amount of residual reflowed resist around the bridges, an ozone treatment at 80$^{\circ}$C for one hour was performed after lift-off.

The final step was to fabricate indium bumps on both chips and bond them together via flip-chip technology. Indium bumps with diameters of 20 - 30\,$\mathrm{\mu m}$ were patterned on both wafers by DWL. Approximately 10\,$\mathrm{\mu m}$-thick indium was grown by thermal evaporation after an in-situ argon ion milling. Resist, and excess indium films were stripped away after soaking in NMP for one day. The carrier and the chip wafers were diced into different sizes and bonded with a bonding force of 150-180\,N in a flip-chip bonder (SET $\mathrm{ACC\mu RA^{TM}\ M}$). Before bonding, all the chips were treated in $\mathrm{H_{2}/N_{2}/He}$ plasma to reduce oxidation layers of the bump surfaces.

\subsection{Measurement setup}

Our measurement setup is shown schematically in Fig.~\ref{fig:wiring}. To reduce the requirement of wirings and electronic resources, we used only one control line for each qubit, combining the qubit drive ($XY$) and flux bias ($Z$) control signals together. Furthermore, we developed a single-sideband (SSB) technology, which was more compact than the traditional IQ mixing scheme and required only half of Digital-to-Analogue Converter (DAC) channels, to generate the qubit drive signal. To obtain a single-tone drive pulse, the baseband waveforms (IF tone) were first generated by a 2\,GSa/s DAC and then mixed with a continuous microwave source (LO tone). Two sidebands with frequencies of $\omega_{\rm LO}\pm \omega_{\rm IF}$ would appear after mixing, with an unwanted LO leakage in the spectrum. On the condition that $\omega_{\rm IF}$ was high enough (for example, $> 200\,\mathrm{MHz}$), we could use a proper bandpass filter to select only one sideband as the control signal. We tested the spectral purity of such SSB technology and could obtain a Spurious Free Dynamic Range (SFDR) of over 50\,dB, which was comparable to or even better than that generated by calibrated IQ mixing method. In addition, as the imaging and leaking LO signals were deeply filtered out, such a scheme required no time-consuming repeated calibrations, which were needed in the traditional IQ mixing scheme.

\begin{figure*}
	\includegraphics[width=0.7\textwidth]{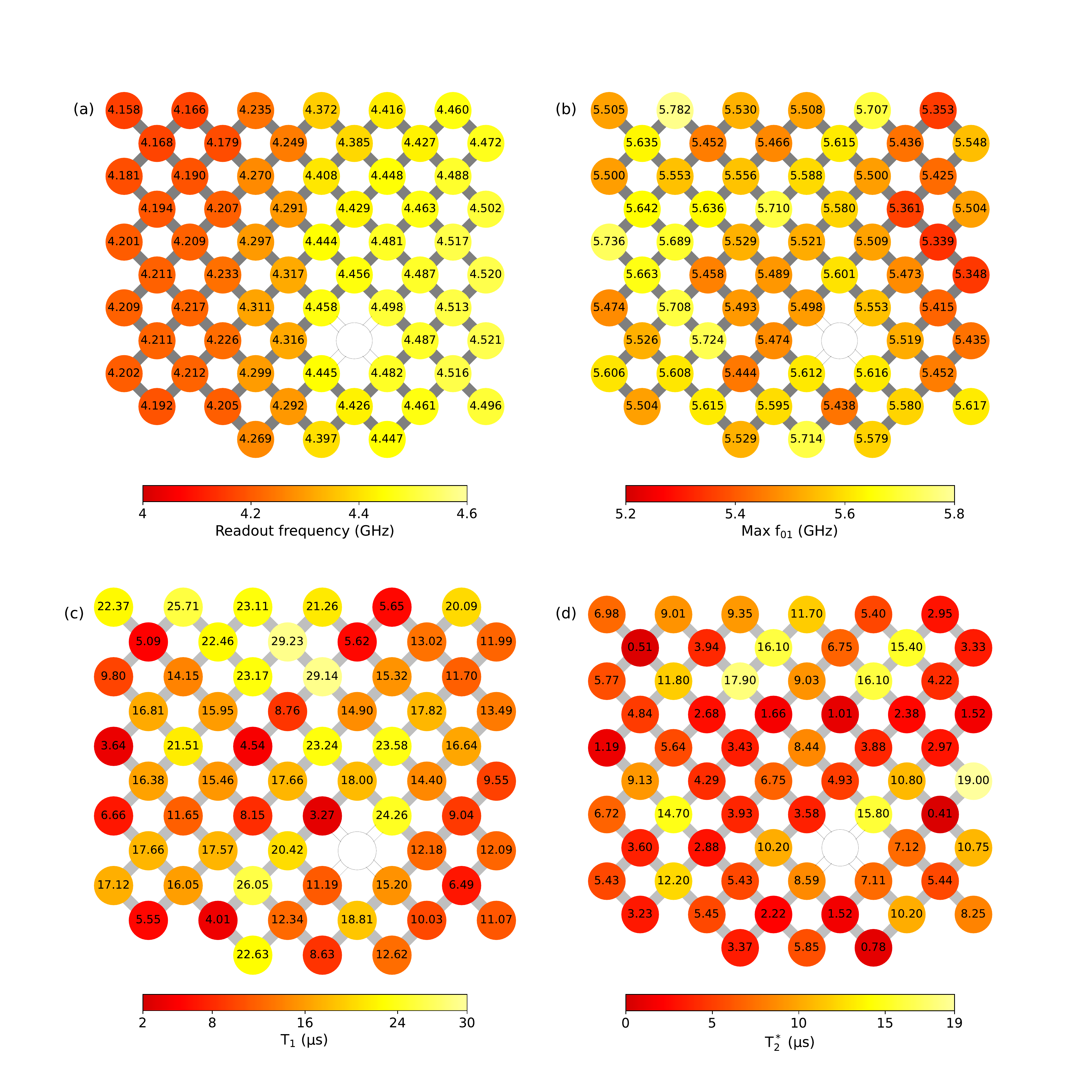}
	\caption{\label{fig:sample} Device parameters. (a) Frequencies of readout resonators. (b) Qubit maximum frequencies. (c) Relaxation time $T_1$ and (d) dephasing time $T_2^{*}$ of qubits at the maximum frequencies. An open circle indicates the qubit that does not work.}
	\label{fig:chipparam}
\end{figure*}

The $Z$ controls of the qubits and couplers were generated directly from 2\,GSa/s DACs, followed by the 1\,GHz lowpass filters. For simultaneous $XY$ and $Z$ controls, the Z pulses were further combined with the $XY$ pulses by using a diplexer. The stimulation signals of the readout resonators were generated by the traditional IQ mixing scheme. All those input signals reached the device at the mixing-chamber (MXC) stage of the dilution refrigerator, with different attenuation at each temperature stage, as shown in Fig.~\ref{fig:wiring}. The filtering configurations at the MXC stage were different for different signals. For the $XYZ$ control of a qubit, the highly attenuated signal was filtered first by a 10\,GHz lowpass filter and an infrared (IR) filter (using Eccosorb$^{\circledR}$ CR124 as the absorber). The CR124 IR filter could attenuate $XY$ control signals for about 20 - 50\,dB (depending on the filter length) with negligible effect on the $Z$ control signals. In addition, it could heavily absorb electromagnetic waves from 10\,GHz to the infrared band. For the $Z$ control of a coupler, it first passed through a 500\,MHz lowpass filter and then a CR110 IR filter (using Eccosorb$^{\circledR}$ CR110 as the absorber). To suppress thermal photon noise, we added 69\,dB attenuation, 10\,GHz low pass filtering, and CR110 IR filtering to each measurement input line in series.

The output of the qubit measurement signal passed through 10\,GHz low pass filter and three isolators to prevent out-band and in-band noises, respectively, from coming down to the chip. It was then pre-amplified by a High-Electron-Mobility Transistor (HEMT) at the 4\,K stage. The signal was further amplified by two microwave amplifiers at room temperature and down-converted to the IF band by an IQ mixer. The converted IQ signals were filtered and amplified and finally digitized by 1\,GSa/s analogue-to-digital converters (ADCs) for demodulation. In order to prevent spurious radiation and flux noise, a light-tight oxygen-free copper shield and a cryoperm shield were added outside the device. In addition, we used another $\mathrm{\mu}$-metal shield (between the 50\,K shield and the vacuum can). The residual magnetic field around the device was measured at room temperature to be less than 20\,nT.

\section{Characterization and calibration}

\subsection{Characterization}

The key parameters of 62 working qubits are depicted in Figs.~\ref{fig:chipparam}(a)-(d), with one that does not work indicated by an open circle. Readout frequencies in Fig.~\ref{fig:chipparam}(a) range from 4.1 to 4.6\,GHz and the maximum frequencies of the qubits in Fig.~\ref{fig:chipparam}(b) from 5.3 to 5.8\,GHz. The relaxation times $T_1$ and dephasing times $T_2^*$ at the maximum frequencies are listed in Figs.~\ref{fig:chipparam}(c) and (d). We find that the qubit coherence times are lower as compared to the values in our previous work~\cite{tapaper}. We address the possible reasons: First, as the circuit complexity increases, a lot of unwanted modes are introduced, which may weakly interact with the qubits and lead to stronger decoherence. Second, although the flipmon design increase the vacuum energy participation ratio, it also increases losses from the metal-air interface. Finally, the combination of $XY$ and $Z$ control signals leads to insufficient filtering, since both the low-frequency $Z$ pulses and high-frequency microwave $XY$ pulses are designed to pass through the control lines. The first two problems can be alleviated by improved design, and the third problem may require a more careful and optimized filtering design.

\begin{figure*}
	\includegraphics[width=0.6\textwidth]{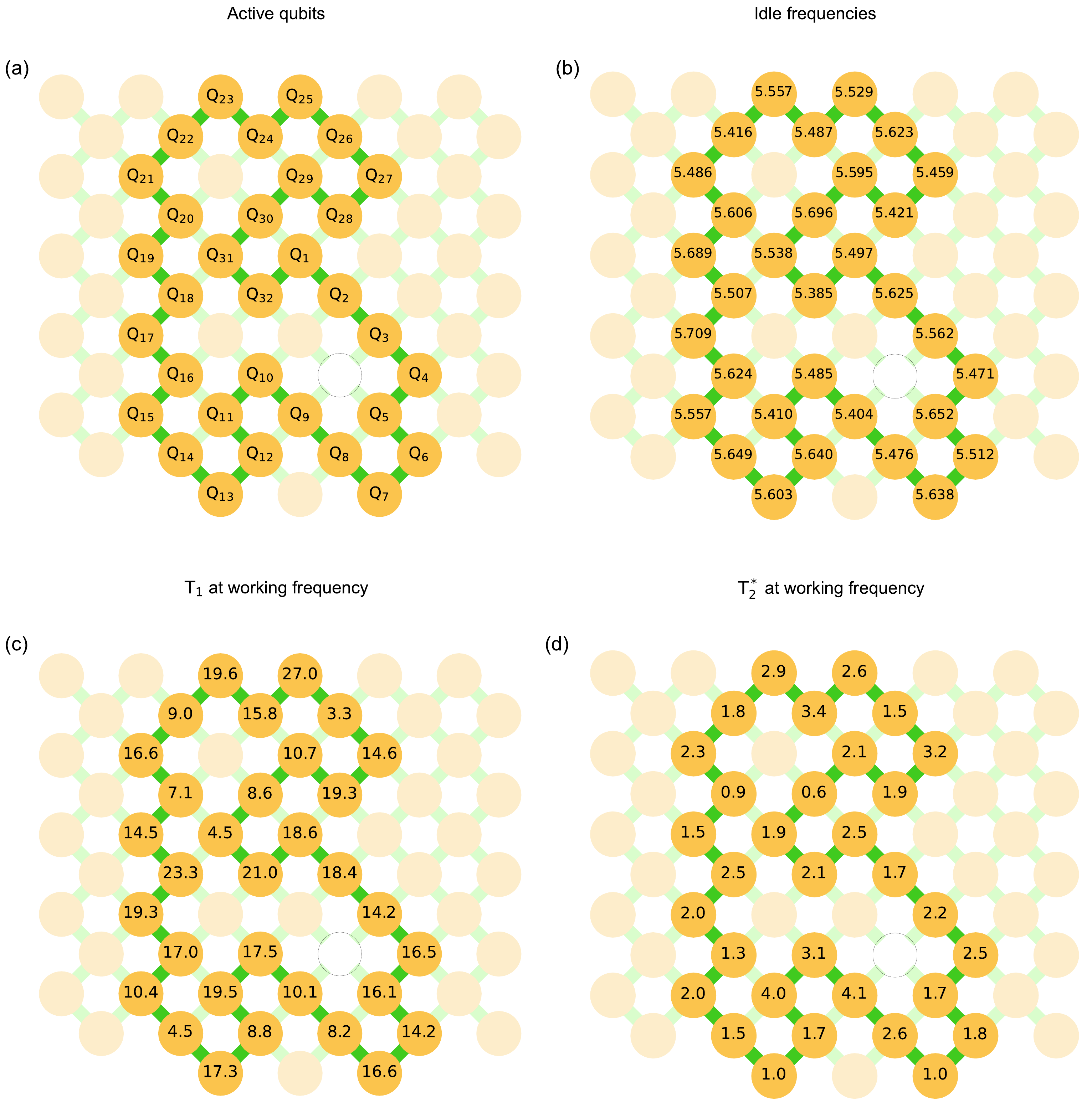}
	\caption{\label{fig:workingpoint} Locations and parameters of 32 active qubits (dark yellow) with 32 active couplers (green) used in the present work. The inactive qubits (light yellow) and couplers (light green) are biased away throughout the experiment. (a) Locations of active qubits $Q_{1}-Q_{32}$. We also label $C_i$ for the coupler between $Q_{i-1}$ and $Q_i$ below, which are not indicated for clarity. (b) Idle frequencies of active qubits.  (c) Relaxation times $T_1$ and (d) dephasing times $T_2^*$ of the active qubits at the working  frequency $\omega_{\rm 0}=2\pi\times5.395\,{\rm GHz}$. An open circle indicates the qubit that does not work.}
\end{figure*}

\begin{figure}[b]
	\includegraphics[width=0.45\textwidth]{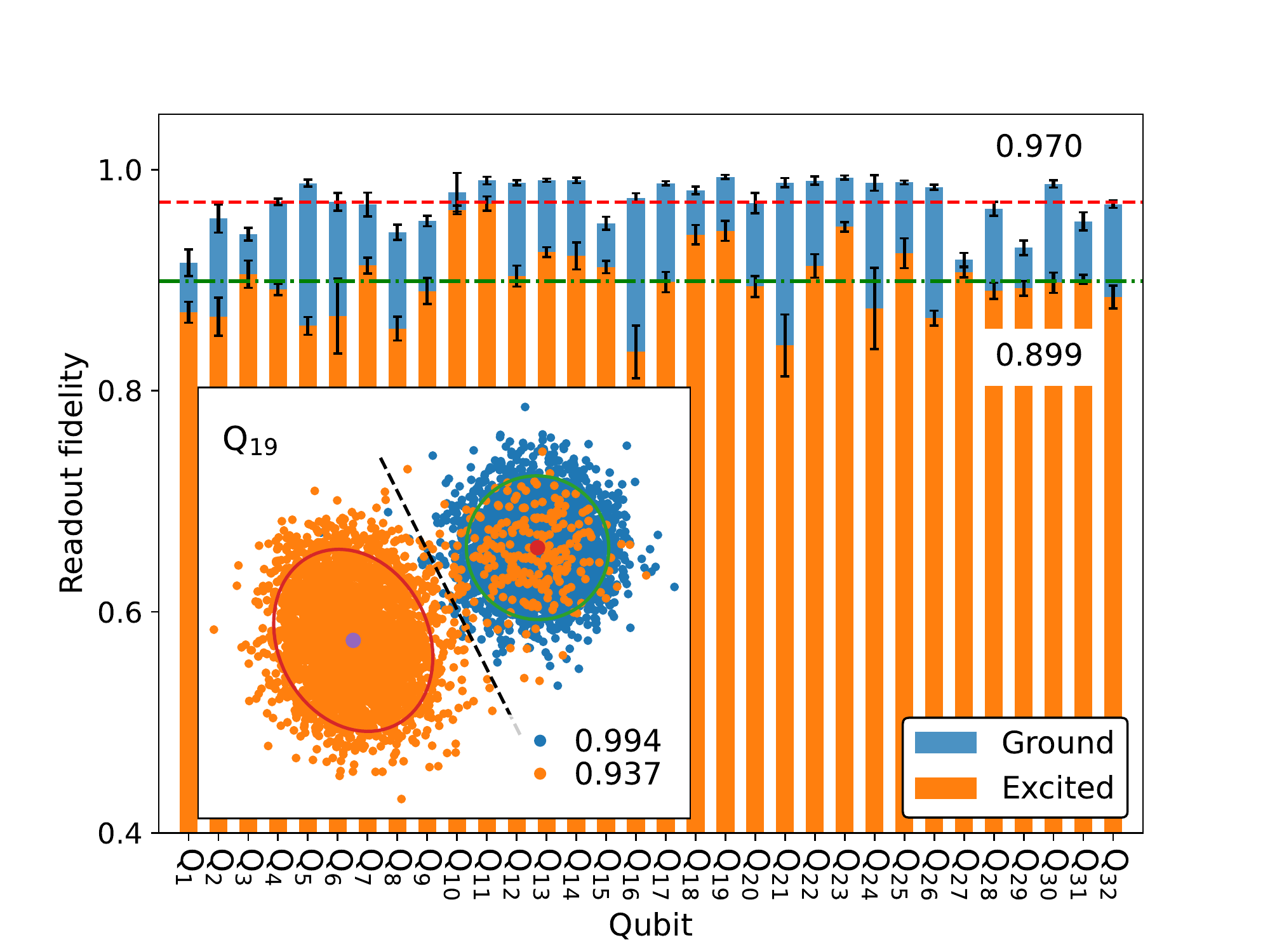}
	\caption{\label{fig:rofidelity} Simultaneous readout fidelities of the active qubits. The average fidelity for the ground (excited) state in blue (orange) bars is 0.97 (0.899), as indicated by the red (green) dashed line. Inset: A typical scattering plot of the readout signal (for $Q_{19}$) in the I/Q plane.}
\end{figure}

Among the 62 working qubits, 32 active qubits, marked as $Q_{1}-Q_{32}$ in Fig.~\ref{fig:workingpoint}(a), were chosen to form a chain for the simulation of the gSSH model in our experiment. All the active qubits were initialized or excited at their idle frequencies shown in Fig.~\ref{fig:workingpoint}(b). They were biased to a common working frequency $\omega_{\rm 0}=2\pi\times5.395\,{\rm GHz}$ for time evolution, during which the inactive qubits were always far-detuned. We measured the $T_1$ and $T_2^*$ of the active qubits at the working frequency, which are shown in Fig.~\ref{fig:workingpoint}(c) and (d).

When multiple qubits were measured simultaneously, the fidelities tended to be lower than those measured individually due to the crosstalk among the control lines. In Fig.~\ref{fig:rofidelity}, we show that the simultaneous readout fidelities of ground and excited states of the active qubits are 97\% and 89.9\% on average, respectively. We mitigated the influence of decoherence on the readout fidelity through the shelving technique~\cite{shelving} on some of the qubits.

\subsection{Crosstalk and distortion corrections}

\begin{figure}[t]
	\includegraphics[width=0.5\textwidth]{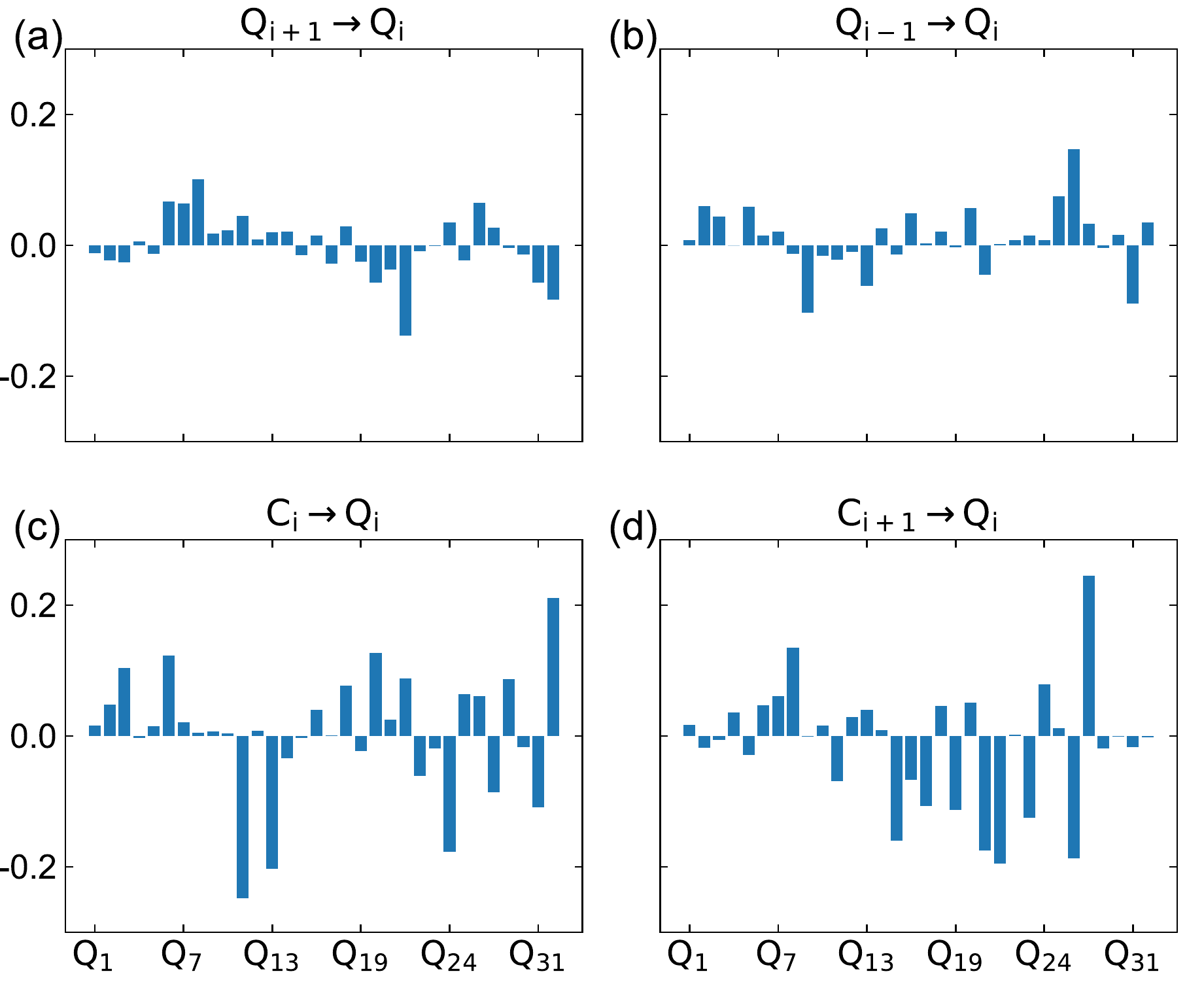}
	\caption{\label{fig:sample} $Z$ crosstalk between adjacent qubits and between adjacent coupler-qubit pairs. (a) Crosstalk between $Q_{i+1}$ and $Q_{i}$. (b) Crosstalk between $Q_{i-1}$ and $Q_{i}$. (c) Crosstalk between $C_{i}$ and $Q_{i}$. (d) Crosstalk between $C_{i+1}$ and $Q_{i}$.}
	\label{fig:crosstalk}
\end{figure}

We reduced the effect of $XY$ crosstalk by carefully choosing the frequency and duration of the drive pulses. $Z$ Crosstalk of nearest qubits and couplers in our chip is shown in Fig.~\ref{fig:crosstalk}. The average $Z$ crosstalk ratio between different flux control lines was less than $0.2\%$. As a result, we ignored $XY$ and $Z$ crosstalk in the experiment.

\begin{figure}[t]
	\includegraphics[width=0.55\textwidth]{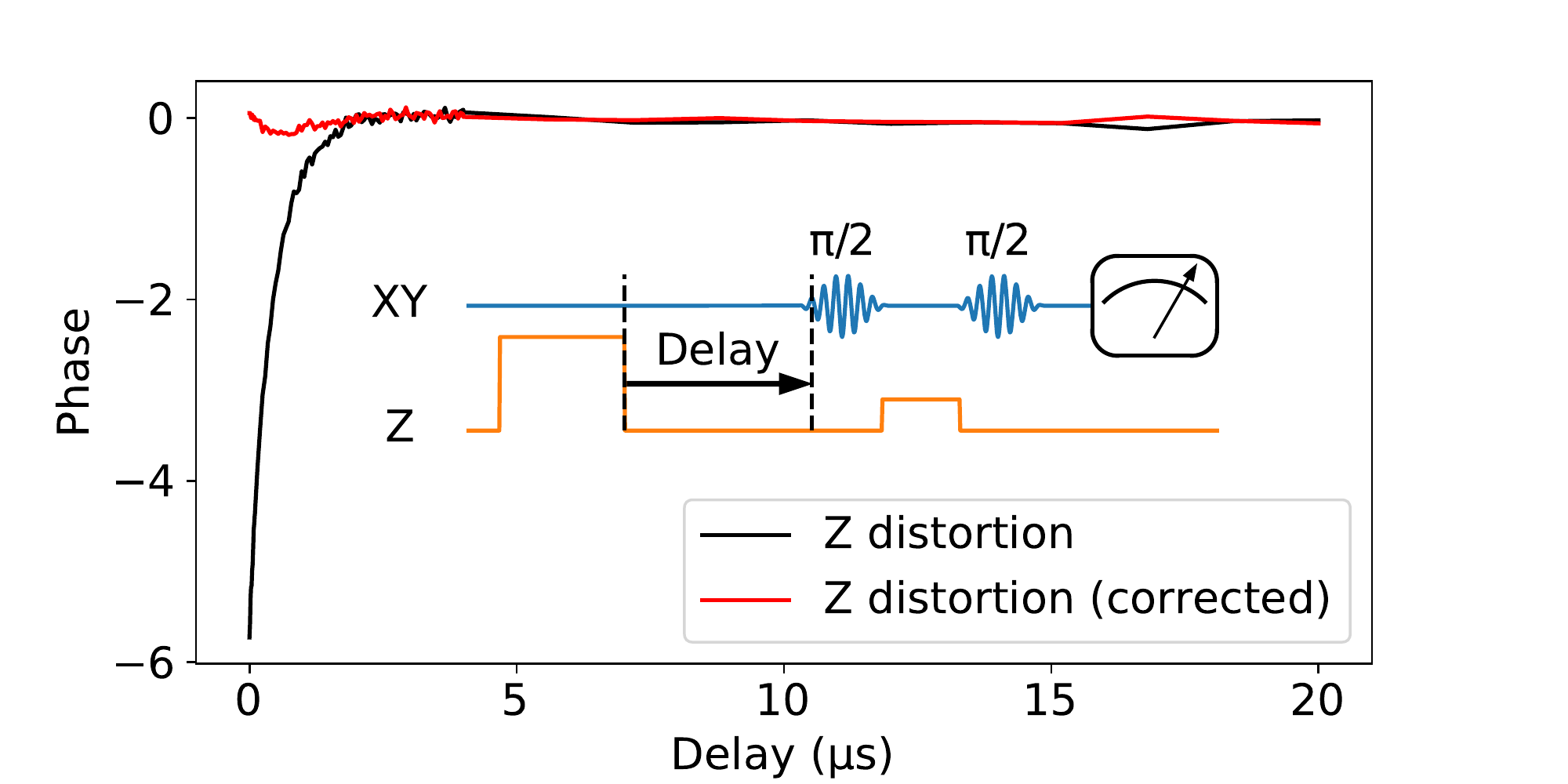}
	\caption{\label{fig:sample}The correction of distorted $Z$-control signal: Experimental results of a distorted square wave (black solid) and a corrected square wave (red solid). Inset: The experimental sequence for the measurement.}
	\label{fig:distortion}
\end{figure}

The nonideal electronics and wirings would make the $Z$-control signal sensed by the qubits severely distorted. In order to obtain accurate control over the qubit, we corrected the distorted signal by the method of deconvolution~\cite{Bao2022fluxonium}. The experimental pulse sequence for measuring the distortion of the $Z$-control signal is shown in the inset of Fig.~\ref{fig:distortion}. First, a square wave was applied on the qubit’s Z-control line, with a large enough amplitude and a long enough pulse duration. Following that, we applied the qubit phase tomography as a distortion detector where a short square wave was inserted between two $\pi/2$ pulses. In order to extract the distortion more accurately, the duration of the $\pi/2$ pulse was set to be short, and the amplitude of the short square wave was carefully selected so that the frequency of the qubit was tuned to a flux-sensitive point. Then, we measured the qubit phase for different delay times between the large square wave and the detector, as shown in Fig.~\ref{fig:distortion} (black line). We calculated the frequency deviation of the qubit according to the measured phase and the duration of the short square wave. Combining with the information of the qubit spectrum, we obtained the trailing amplitude after the large square wave, which represented the response of the system to a step signal. With this we could pre-distort the input signal to correct the distortion. After the correction, the measured phase was observed to be a constant value at different delay times, as shown in Fig.~\ref{fig:distortion} (red line).

\subsection{Timing calibration}

The timing between different control channels was calibrated in the following way. Firstly, we aligned the timing between a single qubit $XY$ and $Z$ controls, and the pulse sequence was shown in Fig.~\ref{fig:calibration}(a). We fixed the time duration between two $\pi$ pulses on the $XY$ control and applied a square wave on the $Z$ control with the same duration. We measured the population of the qubit as a function of the delay between $XY$ and $Z$ control. When the $Z$ pulse was exactly halfway between the two $\pi$ pulses, the population of the qubit should return to zero, as shown in Fig.~\ref{fig:calibration}(d). Secondly, we aligned the Z control timing between the two adjacent qubits by implementing the $i$SWAP-like experiment. As an example, the pulse sequence of $Q_{23}-C_{24}-Q_{24}$ is shown in Fig.~\ref{fig:calibration}(b). Note that the square wave duration of the coupler is set larger. When the two $Z$ pulses were aligned correctly, the population exchange between the two qubits reached the maximum, which is shown in Fig.~\ref{fig:calibration}(e). Finally, the alignment of the Z pulses of the adjacent coupler and qubit was carried out with a similar method, as shown in Figs.~\ref{fig:calibration}(c) and (f).

\subsection{Coupling strength}

\begin{figure*}[t]
	\includegraphics[width=0.72\textwidth]{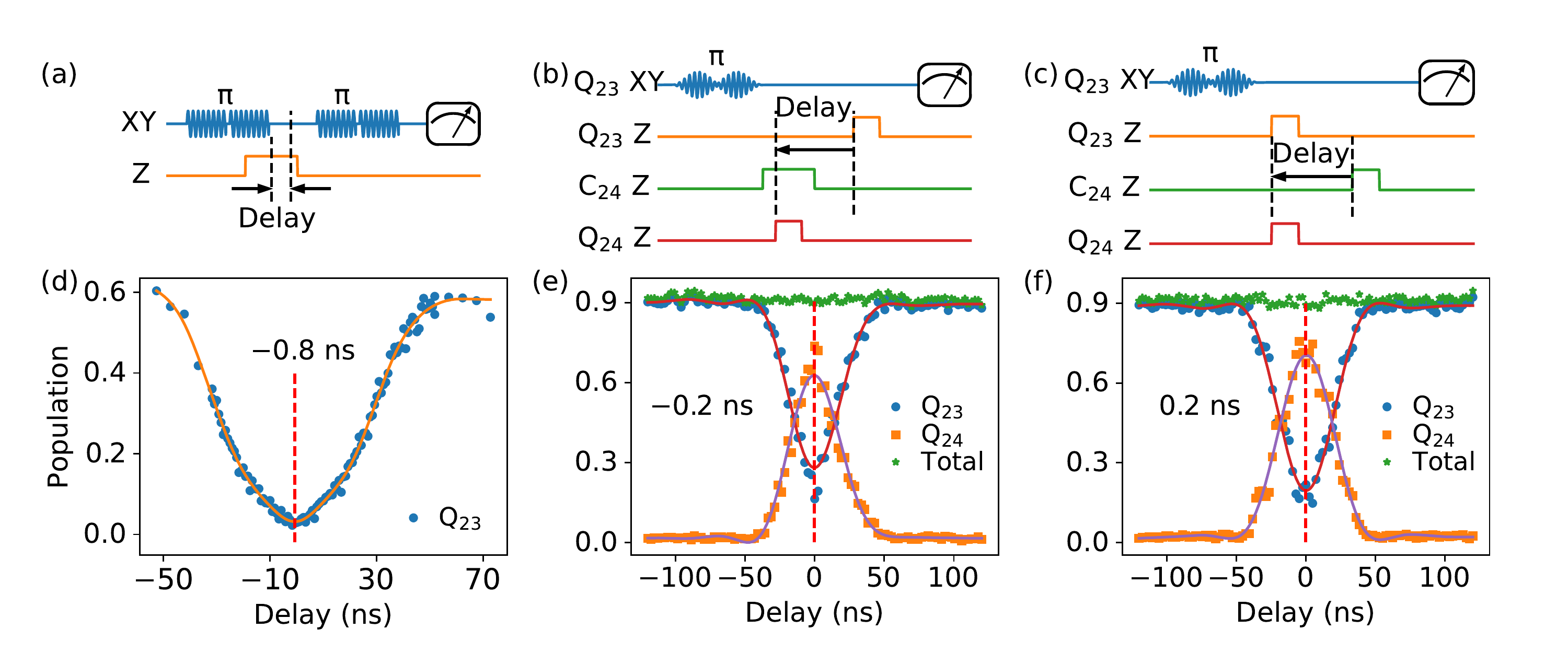}
	\caption{Timing calibration. (a), (b), and (c) show the pulse sequences for the timing alignments between qubit’s $XY$ and $Z$ controls, between $Z$ controls of two adjacent qubits, and between $Z$ controls of a qubit and its nearest coupler, respectively. (d), (e), and (f) show the corresponding typical experimental results (dots). The solid lines are the experimental data filtered by a fifth-order low-pass Butterworth filter.}
	\label{fig:calibration}
\vspace{8mm}
	\includegraphics[width=0.75\textwidth]{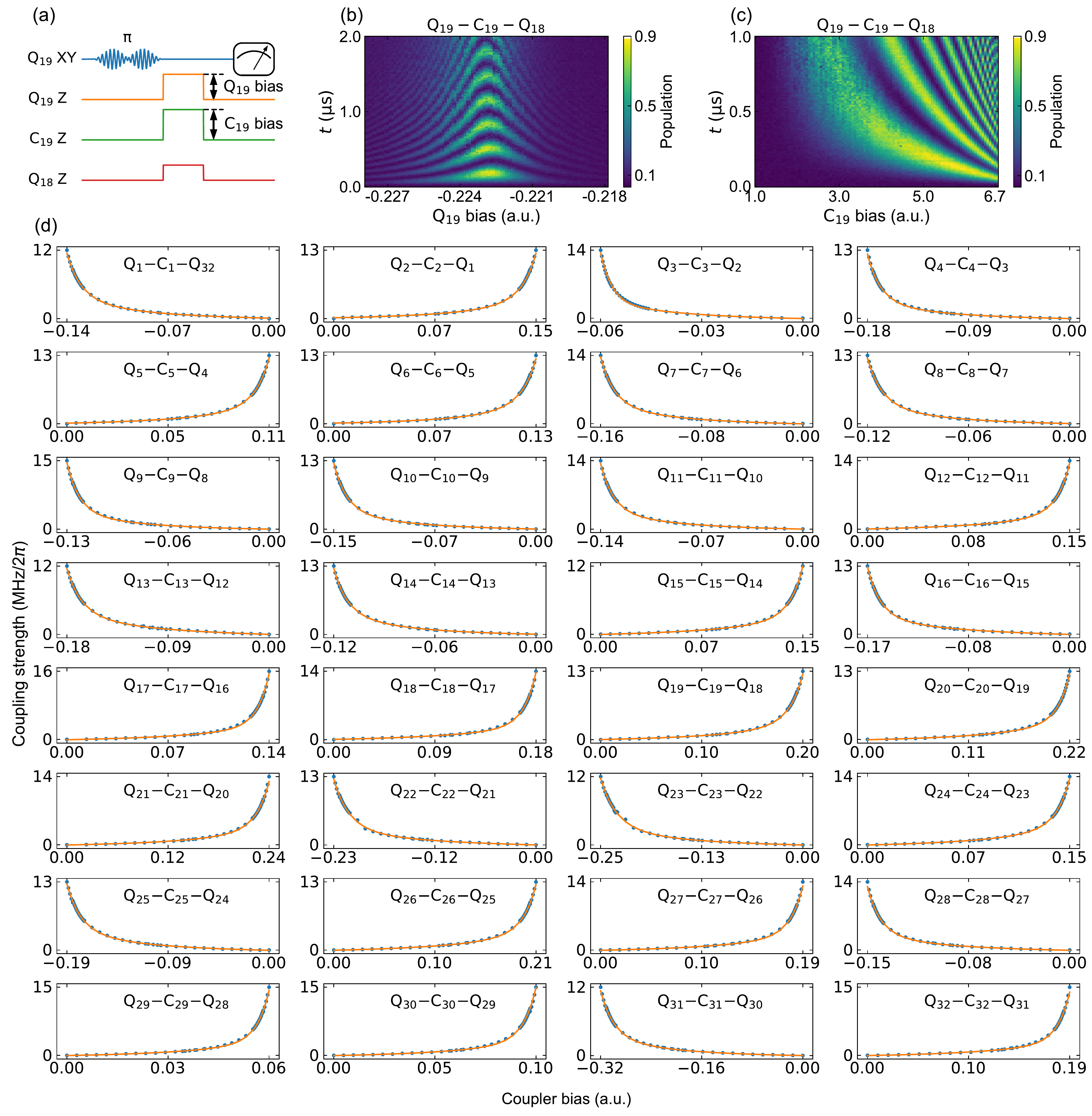} 
	\caption{Measurements of tunable coupling strength between nearest-neighbor qubit pairs. (a) \label{fig:s9a} The sequence of $i$SWAP experiment. (b) The typical $i$SWAP oscillation between two qubits ($Q_{19}$ and $Q_{18}$), when the coupling strength ($C_{19}$) is $\mathrm{2\pi\times1.5\,MHz}$. (c) The oscillation frequency as tuned by the flux bias of the $C_{19}$. (d) The coupling strength of 32 qubit-coupler-qubit triples as a function of coupler bias. The results are parameterized by training a polynomial model using ridge regression. Dots (lines) are the experimentally extracted (parameterized) coupling strength.}
	\label{fig:coupling}
\end{figure*}

In the experiment, it was crucial to accurately determine the effective coupling strength between each pair of nearest-neighbor qubits. To do so, we used the pulse sequence shown in Fig.~\ref{fig:coupling}(a), taking a subsystem ($Q_{18}-C_{19}-Q_{19}$) as an example. We first decoupled the surrounding qubits and couplers, and prepared the initial state as $\ket{100}$ by applying a $\pi$ pulse on $Q_{18}$. Then, we measured the excitation probability of $Q_{19}$ as a function of the evolution time $t$ and the flux bias of $Q_{19}$, which is shown in Fig.~\ref{fig:coupling}(b). It is observed that the excitation swaps between the two qubits. The typical chevron pattern where swap frequency is minimum indicates the resonance point at which the frequencies of two qubits are equal. We fixed the flux biases of qubits to keep them resonant and then measured the excitation probability of $Q_{19}$ as a function of the flux bias of $C_{19}$, shown in Fig.~\ref{fig:coupling}(c). We extracted the effective coupling strength by fitting the data along each vertical line. The obtained values varied in the range of $2\pi\times\left[-14, 0\right]$ MHz. Because of the limited evolution time up to 2\,$\mu$s, the extracted coupling strength from -14\,MHz to -0.25\,MHz was inferred. Continuous parameterization of the effective coupling strength was important to quickly simulate the target Hamiltonians. In our experiment, the extracted coupling strength from -14\,MHz to -0.25\,MHz as a function of the coupler flux bias was used to train a polynomial model by ridge regression~\cite{pedregosa2011scikitlearn}, which predicted the values for the flux bias corresponding to given coupling strengths in target Hamiltonians. Besides, the switch-off point with vanishing coupling strength was calibrated with a much longer pulse sequence, which was comparable to the energy relaxation time $T_1$ of the involved qubits. So we continuously parameterized the coupling strength by linear interpolation from -0.25\,MHz to 0\,MHz. Figure~\ref{fig:coupling}(d) shows a total of 32 parameterized coupling strengths between the nearest-neighbor qubit pairs.

\begin{figure}
	\centering
	\includegraphics[width=0.52\textwidth]{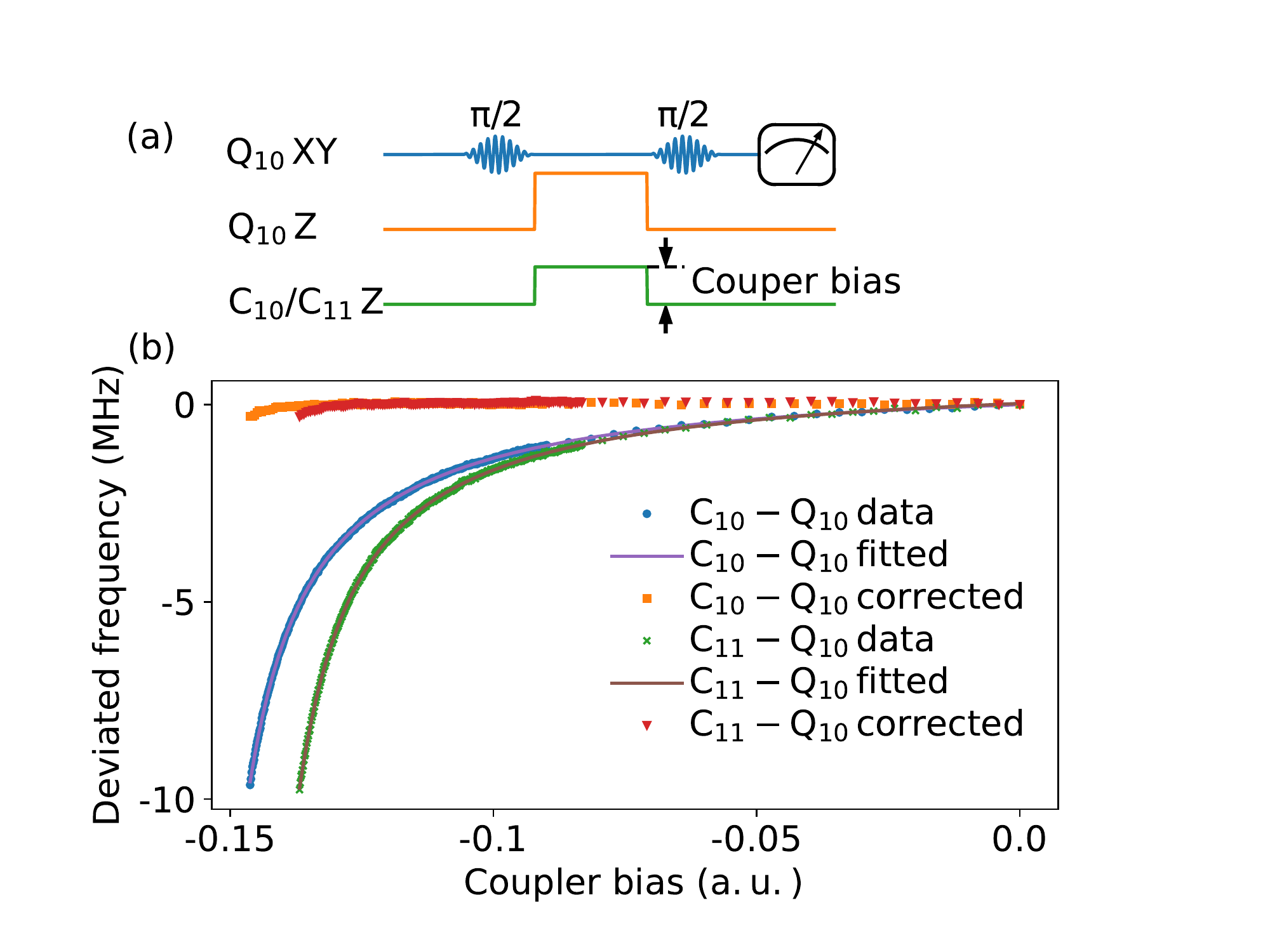} 
	\caption{Calibration of the effect of the coupler bias on the qubit frequency. (a) Experimental pulse sequence. (b) The blue circles (green crosses) show the experimental frequency shifts on $\mathrm{Q_{10}}$ when the bias of coupler $\mathrm{C_{10}}$ ($\mathrm{C_{11}}$) changes and the coupling strength decreases. The orange squares ($\mathrm{C_{10}}$) and red triangles ($\mathrm{C_{11}}$) are experimental data after calibration. The solid lines are the fitted results.}
	\label{fig:AC stark}
\end{figure}

\begin{figure}[b]
\centering
\includegraphics[width=0.95\linewidth]{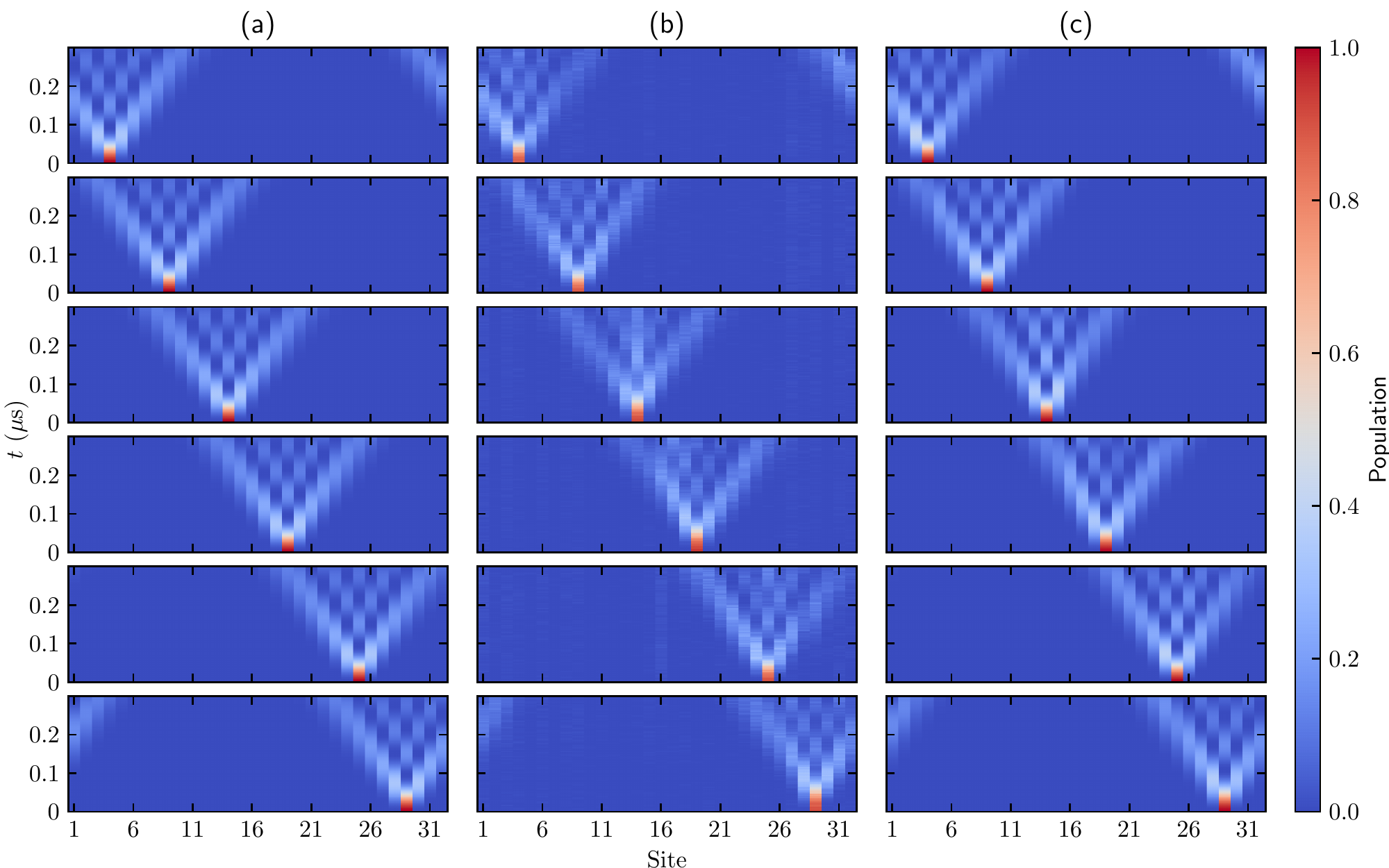}
\caption{Single excitation quantum walk in the 32-qubit chain with all the nearest-neighbor coupling strength set to 2\,MHz. (a) Theoretical and (b) experimental results. (c) Fitted results by optimizing the Hamiltonian to approximate the actual evolution in (b). From top to bottom rows, qubits Q$_{4}$, Q$_{9}$, Q$_{14}$, Q$_{19}$, Q$_{25}$, and Q$_{29}$ are initially excited, respectively. See text for more details.}
\label{fig:light_cone2}
\end{figure}

\subsection{Frequency alignment}

We biased all qubits to the working point with frequency $\omega_{\rm 0}=2\pi\times5.395\,{\rm GHz}$ for the quantum walk experiment. For this we fixed one qubit at the working point and applied an $i$SWAP sequence (see Fig.~\ref{fig:s9a}) between a qubit and its nearest neighbor qubit. During the $i$SWAP experiment, the coupling strength between the two qubits was set to be 0.25\,MHz, and the rest qubits were decoupled from these two. We repeated this two-qubit frequency alignment process sequentially along the qubit chain. As this procedure traversed the entire 32-qubit chain, we found the detuning between first qubit and last qubit to be very small. Due to inherent many-body nature of the system, when all the qubits and couplers were set to the points we characterized above in parallel, the alignment might not be perfect. Figure~\ref{fig:AC stark} shows that the coupling of a qubit to a coupler can deviate the qubit's frequency from the working point for about -10\,MHz. Here, we parameterized the deviated qubit frequency as a function of the coupler bias by a polynomial model with ridge regression. We compensated for this deviation with an extra qubit bias. With this we see that the qubit frequency becoming nearly constant during sweeping the coupler bias. Due to the limited compensation accuracy of this method, we only compensated the qubit bias within the range of coupling strength from -9.5\,MHz to -0.25\,MHz.

\subsection{Target coupling and parameterized coupling}

\begin{figure}
	\centering
	\includegraphics[width=0.45\textwidth]{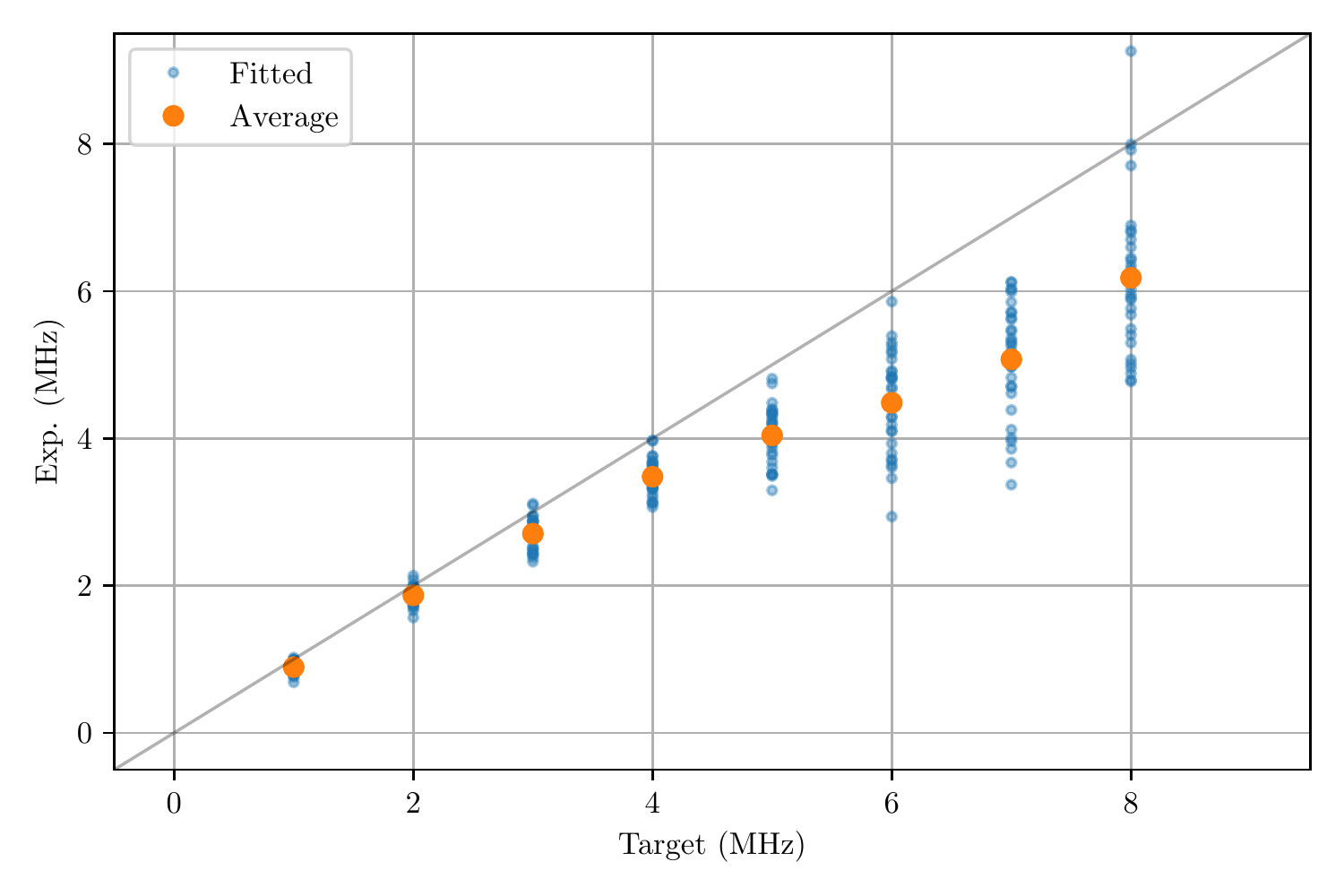} 
	\caption{Comparison between parameterized coupling strength and target coupling strength. Blue dots are the approximate coupling strength from the optimization of the coupling strength of 32 qubit pairs in the Hamiltonian. Yellow dots are their average.}
	\label{fig:light_cone}
\end{figure}

Although we were able to accurately parameterize the coupling strength between each nearest-neighbor qubit pair when all other qubits are decoupled, it was still challenging to parameterize the coupling strength when all qubits and couplers are activated, due to the intrinsic many-body nature of the device. Therefore, we evaluated the extension of our pair-wise coupling strength parameterization to the whole qubit chain with quantum walk experiments. We implemented the 32-qubit single-excitation quantum walk using the target Hamiltonian, where all couplings were set to be the same target strength. For example, we performed quantum walk experiments with six different initial states, with the target coupling strength set to be 2\,MHz. The results of up to 300\,ns evolution were shown in Fig.~\ref{fig:light_cone2}. The left column shows the theoretical simulation results of the quantum walks using the target Hamiltonian. The middle column shows the experimental results of the quantum walks of the parameterized Hamiltonian. To accurately extract the actual coupling strength of each qubit pair, we optimized the Hamiltonian to match the experimental results, where the coupling strengths were the fitting parameters. The results are shown in the right column. We can see a good consistency of the target coupling to parameterized coupling. Repeating this procedure with different target coupling strengths, we were able to check the consistency of the experimentally extracted coupling strength with the target coupling strength, shown in Fig.~\ref{fig:light_cone}. It can be seen that the pair-wise coupling strengths are well in control when the target coupling strength is below 4\,MHz.

\begin{figure*}
	\centering
	\includegraphics[width=\textwidth]{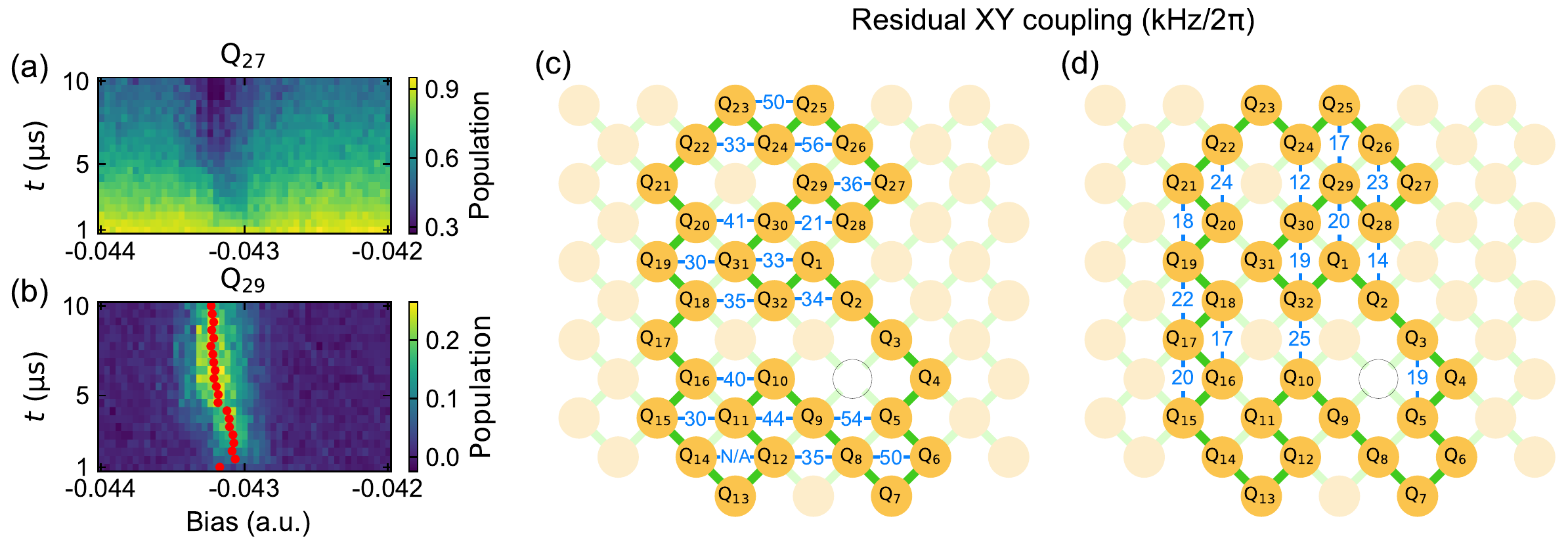} 
	\caption{Demonstration of the residual $XY$ coupling strength. (a) and (b) show the typical characteristic of swapping between $Q_{27}$ and $Q_{29}$. The red dots in (b) illustrate the evolution of population of $Q_{29}$, which can be fitted to obtain the residual coupling strength. (c) and (d) show the residual coupling strength between two diagonal next-nearest neighbor qubits. The average residual coupling strength is $2\pi\times30$ kHz. N/A in (c) indicates a coupling strength experimentally unobservable due to its small value.}
	\label{fig:Residual couplings}
\end{figure*}
	
\subsection{Residual coupling}

For the superconducting quantum processor consisting of qubits arranged in 2-dimensional square lattice with a coupler between each of the two nearest-neighbor qubit pairs, one can only turn on and off the coupling between the nearest-neighbor qubit pairs. The residual coupling between diagonal next-nearest neighbor (NNN) qubits will have a negative impact on the experimental results. However, for the present device in which flipmon-type qubits are used~\cite{flipmonpaper}, the residual coupling between NNN qubits are very small. To see this, we applied the $i$SWAP experiment to NNN qubit pairs to evaluate the residual coupling strength. The population swapping results are shown in Figs.~\ref{fig:Residual couplings} (a) and (b) taking $Q_{27}$ and $Q_{29}$ as an example. To extract the small coupling strength was difficult owing to the limited coherence time and residual Z distortion. We selected the maximum swapping population among biases to mitigate the distortion and fitted it as a function of evolution time using QuTiP~\cite{Johansson_2012}. Here we fixed the $T_1$ and $T_2^{*}$ of two qubits measured at the working frequency and the coupling strength served as the fitted parameter. Figures~\ref{fig:Residual couplings} (c) and (d) are the measured NNN coupling strength for the 32-qubit chain, which ranges from values experimentally unobservable to a maximum of 56 kHz. The average residual coupling strength of the device is about $2\pi\times30$ kHz, corresponding to a swapping period of 17\,$\mu$s. In our quantum walk experiment in the main text, the evolution time is set to be 1.5\,$\mu$s so that the residual coupling does not pose a significant impact on our experiment.


\section{Quantum simulation experiment}

\subsection{Generalized SSH model}

The target model considered in this experiment is the generalized SSH (gSSH) model with the intra-cell tunneling strength modulated by a quasiperiodic disorder. Here we recall the Hamiltonian as follows:
\begin{eqnarray}
\hat H_{\rm gSSH}/\hbar=\sum_{n=1}^{N_c}m_n\hat a_{n}^\dag\hat b_{n}+g\sum_{n=1}^{N_c-1}\hat a_{n+1}^\dag\hat b_{n}+{\rm h.c.},\label{eq:hamil_gSSH}
\end{eqnarray}
with homogeneous inter-cell coupling $g$ and intra-cell tunneling strength of
\begin{eqnarray}
m_n=m+W\cos\left(2\pi\beta n+\delta\right) \label{eq:m_n}
\end{eqnarray}
with $\beta=\frac{\sqrt{5}-1}{2}$.

\subsection{Quantum walks}

\begin{figure*}[t]
	\includegraphics[width=0.85\linewidth]{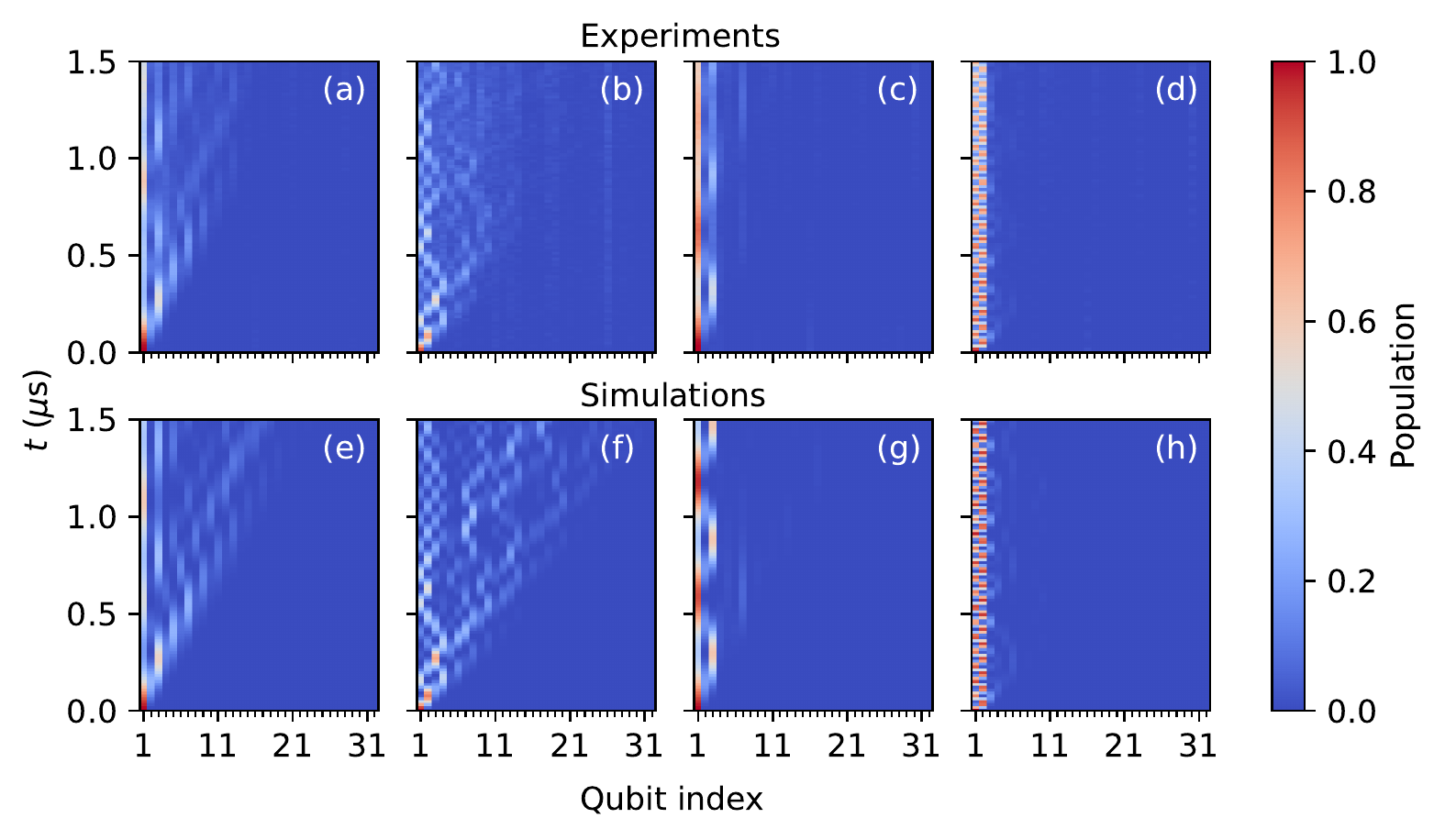} 
	\caption{Time-dependent particle density distribution for a single particle initially excited at the edge (Q$_1$), after quenching the system to (a) the topological extended phase with $(m/g,~W/g)=(0.1,~0.6)$, (b) the trivial extended phase with $(m/g,~W/g)=(0.5,~2.0)$, (c) the topological localized phase with $(m/g,~W/g)=(1.0,~0.5)$ and (d) the trivial localized phase with $(m/g,~W/g)=(3.0,~1.25)$. The qubit site index ranges from 1 to 32. (e)-(f) Corresponding numerical results obtained from evolving the Schr\"{o}dinger equation.}
	\label{fig:quantum_walks}
\end{figure*}

The topological and localization properties of the system can be visually shown from the quantum walk experiment. To this end, we target four Hamiltonians with the parameter sets ($W/g$, $m/g$) of (0.1, 0.6), (0.5, 2.0), (1.0, 0.5), and (3.0, 1.25), corresponding to topologically extended, trivial extended, topologically localized, and trivial localized states, respectively. We then prepare an initial state with a single excitation at the edge. The experimental results of particle density evolution, obtained from projective measurement after turning on the target Hamiltonian and evolving the system for a time period $t$, are shown in Fig.~\ref{fig:quantum_walks}. The difference between extended and localized phases with or without topological order is clearly observed. As a topological feature, the edge excitations for topological phases remain largely at the edge throughout the evolution no matter the bulk states are localized or extended, while in the trivial phases the particle density will remain at the edge in the localized phase but will spread to all qubits in the extended phase. Moreover, the edge excitations in topological phases mainly couple and spread slightly to nearby sites in the same sublattice.

The experimental results in Figs.~\ref{fig:quantum_walks}(a)-(d) are obtained by mitigating the readout error under the assumption of uncorrelated readout noise~\cite{nation2021scalable} and post-selecting the mitigated probability density distribution in the single-excitation subspace, which show a fair agreement with the corresponding numerical results in Figs.~\ref{fig:quantum_walks}(e)-(h). However, there are some visible discrepancies. For example, the dynamical wave function is less extended in experiment compared to that in numerical simulation for the extended phases. This stems from the residual frequency detuning with respect to the global working frequency, to which the extended states are more vulnerable. Besides, spurious vertical lines can be seen in Fig.~\ref{fig:quantum_walks}(b) due to the drifting of readout parameters.

\subsection{Dynamical observables}

The topology of the 1D gSSH model can be characterized by the winding number $N_{\rm w}$, an integer-valued topological invariant originally defined in the momentum space, with $N_{\rm w}=1$ indicating nontrivial topology. When disorder is present, the topology can be revealed by the real-space representation~\cite{mondragon2014topological}. To see this, we write the winding number $N_{\rm w}$ for a 1-D system as:
\begin{equation}
N_{\rm w} = \frac{1}{L_B}{\rm Tr}_B\left(\hat \Gamma \hat Q\left[\hat Q, \hat X\right]\right),\label{eq:winding_number}
\end{equation}
with ${\rm Tr}_B$ being the trace over the bulk regime of length $L_B$, $\hat\Gamma=\sum_n\left(\hat a_n^\dag\hat a_n-\hat b_n^\dag\hat b_n\right)$, and $\hat X=\sum_n n\left(\hat a_n^\dag\hat a_n+\hat b_n^\dag\hat b_n\right)$. The operator $\hat Q$ is obtained by flattening the energy spectrum of $\hat H_{\rm gSSH}$, i.e., $\hat Q=\sum_n{\rm sgn}\left(E_n\right)\ket{\phi_n}\bra{\phi_n}$, where $\ket{\phi_n}$ is the eigenstate of $\hat H_{\rm gSSH}$ with the eigenenergy of $E_n$ and ${\rm sgn}\left(\cdot\right)$ is the sign function.

It is proposed and demonstrated in Ref.~\cite{cardano2017detection} that the winding number can be measured in experiment by the mean chiral displacement. Here we denote the mean chiral displacement for a specific quantum-walk experiment as
\begin{eqnarray}
	C_t=\frac{2}{t}\int_0^t\left\langle\psi\left(t'\right)\left|\hat\Gamma\hat X\right|\psi\left(t'\right)\right\rangle dt'.
\end{eqnarray}
where the quantum-walk dynamics is governed by $\left|\psi\left(t\right)\right\rangle=\exp\left(-i \hat H_{\rm gSSH}t/\hbar\right)\left|l_0\right\rangle$ with the initial state $\ket{l_0}$ being the Fock state with only the $l_0$-th site excited. The winding number $N_{\rm w}$ can be obtained as $N_{\rm w}=\lim_{t\rightarrow\infty}\overline{C}_t$, with an overline denoting averaging over different quantum-walk realizations.

To characterize the localization properties, we use the spectral average of the inverse participation ratio~\cite{evers2008anderson}, defined as
\begin{eqnarray}
\overline{\rm IPR}\equiv L^{-1}\sum_n{\rm IPR}_n,~{\rm with}~{\rm IPR}_n=\sum_{l=0}^{L-1}\left|\overlap{l}{\phi_n}\right|^4, \label{eq:average_IPR}
\end{eqnarray}
where $\ket{l}$ is the Fock state with only the $l$-th site excited and $\ket{\phi_n}$ the $n$-th eigenstate of $\hat H_{\rm gSSH}$. In the thermodynamic limit, $\overline{\rm IPR}$ tends to zero for systems in the extended phases, while it remains finite otherwise. 

In this experiment, the average IPR is measured by the mean survival probability, defined as
\begin{eqnarray}
S_t(l_0)=\frac{1}{t}\int_0^t\left|\overlap{l_0}{\psi\left(t'\right)}\right|^2dt',
\end{eqnarray}
here $\ket{l_0}$ is the initial state of the quantum-walk experiment. It can be seen that $\overline{\rm IPR}=\lim_{t\rightarrow\infty}\overline{S}_t$, with $\overline{S}_t=\frac{1}{L}\sum_{l_0}S_t(l_0)$. Using the completeness relation of the eigenstates of the gSSH model, we can calculate $S_\tau(l_0)$ as
\begin{widetext}
\begin{eqnarray}
S_t(l_0)&=&\frac{1}{t}\int_0^t\sum_{n,m}\overlap{\psi\left(t'\right)}{\phi_n}\overlap{\phi_n}{l_0}\overlap{l_0}{\phi_m}\overlap{\phi_m}{\psi\left(t'\right)}dt'\\
&=&\sum_n\left|\overlap{l_0}{\phi_n}\right|^4+\sum_{n,m\neq n}\frac{1}{i\omega_{n,m}t}\left(e^{i\omega_{n,m}t}-1\right)\left|\overlap{l_0}{\phi_n}\right|^2\left|\overlap{l_0}{\phi_m}\right|^2,\nonumber
\end{eqnarray}
\end{widetext}
with $\omega_{n,m}=\hbar^{-1}\left(E_n-E_m\right)$ and $\ket{\phi_{n}}$ being the eigenstate of $\hat H_{\rm gSSH}$ with the eigenenergy $E_{n}$. It is clear that the second term vanishes as $t\rightarrow\infty$ and thus the average of $S_\infty\left(l_0\right)$ over the initial excitation position $l_0$ in the long-time limit gives $\overline{\rm IPR}$. 

\begin{figure}[b]
\includegraphics[width=\linewidth]{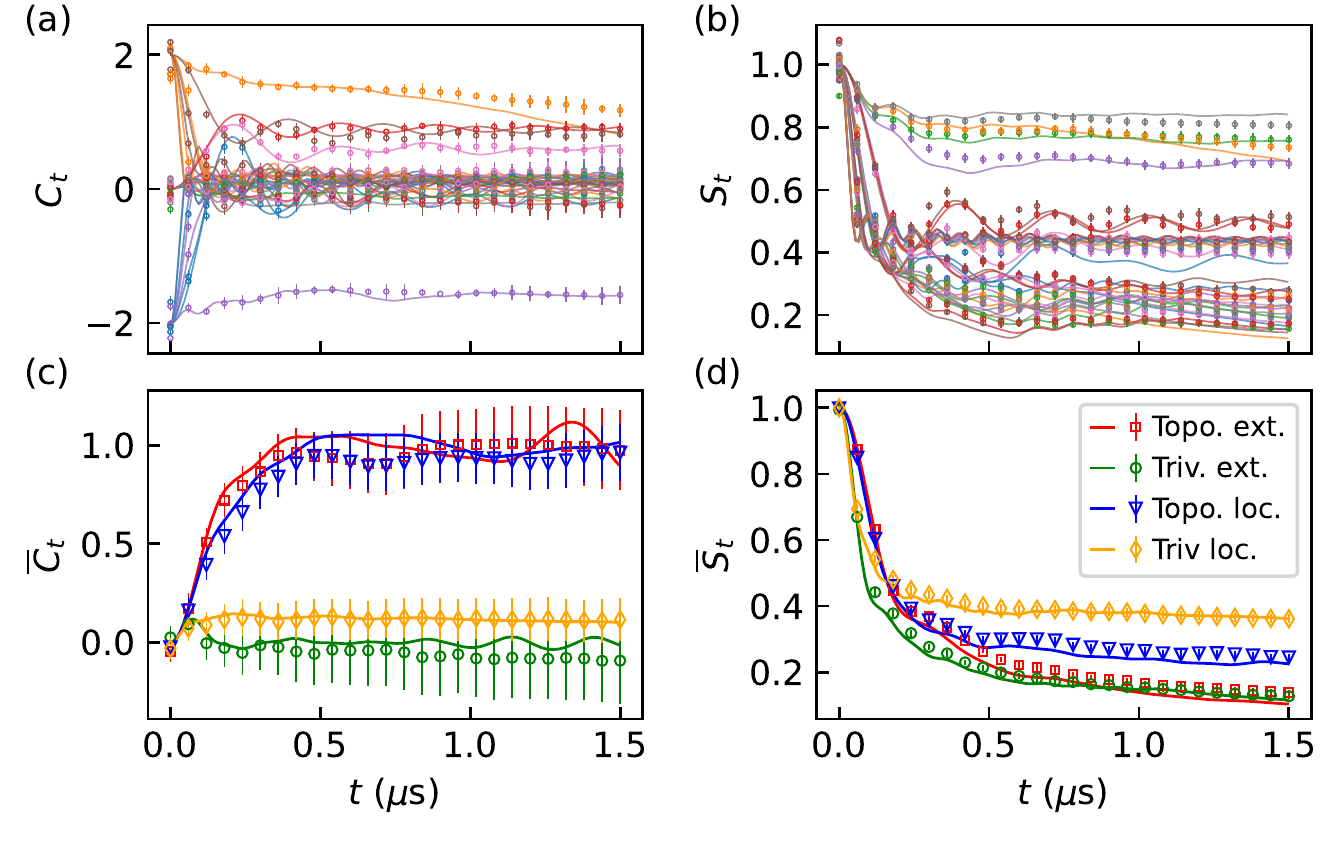}
\caption{Dynamical observables for the topological and localization properties. (a) Time-averaged chiral displacement and (b) survival probability of 32 different realizations for $W/g=3$ and $m/g=1.25$. The realizations are generated by setting $\delta\in\left[0,\ldots,7\right]\times\pi/4$ and placing the initial excitation at sites $l_0\in\left\{15, 16, 17, 18\right\}$. Here symbols and lines are the experimental and numerical results, respectively. For the experimental results, the error bars represent the standard deviation of the mean arising from the sampling error in the projective measurement with 1024 repetitions. (c) Average chiral displacement and (d) average survival probability as a function of the evolution time $t$ for the gSSH model in the topological extended, trivial extended, topological localized and trivial localized phases. The parameters $W$ and $m$ are the same as those used in the quantum-walk experiment in Fig.~\ref{fig:quantum_walks}.}
\label{fig:dynamical_observables}
\end{figure}

To reduce the impact of the finite-size effect, we construct 8 disorder realizations and 4 initial single-excitation states in the bulk for a given target Hamiltonian. Figures~\ref{fig:dynamical_observables}(a) and (b) show the dynamical observables of $C_t$ and $S_t$ for the 8$\times$4 quantum-walk instances in the trivial localized phase with $W/g$ = 3 and $m/g$ = 1.25, where the instance-averaged $\overline{C}_t$ and $\overline{S}_t$ are shown as squares in Figs.~\ref{fig:dynamical_observables}(c) and (d). Although the evolutions of $C_t$ and $S_t$ depend on the specific disorder realization and initial state, $\overline{C}_t$ and $\overline{S}_t$, after taking average over the 32 quantum-walk instances, converge to $\nu$ and $\overline{\rm IPR}$, respectively. Besides, the mean chiral displacements and  survival probabilities for the systems in the other three phases, with the specific parameters indicated in the caption of Fig.~\ref{fig:quantum_walks}, are also present in Figs.~\ref{fig:dynamical_observables}(c) and (d).


We have seen that in the thermodynamic limit and with infinite evolution time, the mean chiral displacement and survival probability would converge to the winding number and spectrally averaged IPR. In experiment, the qubit number of our superconducting simulator are far from reaching the thermodynamic limit, and the evolution time of quantum-walks is also limited by the coherence times of the qubits. In the following, we assess the afficacy of the experimental protocol from three aspects: (i) the number of disorder realizations; (ii) the finite size of the experimental system; (iii) the finite evolution time.

\subsection{Effect of finite disorder realizations}

\begin{figure}[b]
	\centering
	\includegraphics[width=\linewidth]{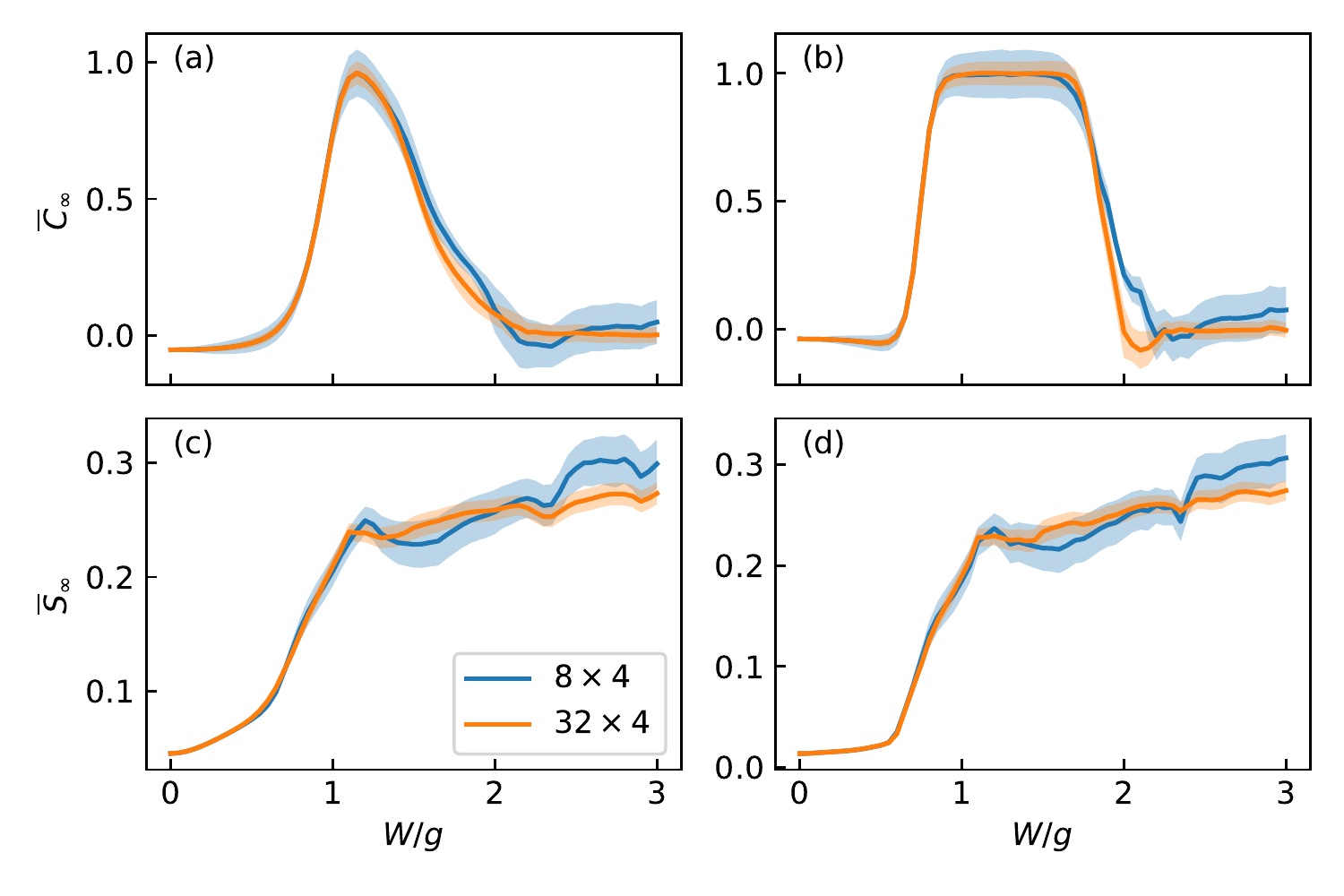}
	\caption{Numerical results for the mean chiral displacement (a, b) and survival probability (c, d) averaged over different numbers of disorder realizations for the gSSH model with $m/g=1.1$. The system size is $N_c=16$ for (a) and (c), which is the same as the experiments, while in (b) and (d) we present results for a larger system size $N_c=55$. The lines are labeled by $M_{\rm dis}\times M_{\rm ini}$, with $M_{\rm dis}$ and $M_{\rm ini}$ denoting the number of disorder realizations and the number of initial states, respectively. The shaded regions indicate the error of the mean.}
	\label{fig:disorder_realization}
\end{figure}

The numerical results in Fig.~\ref{fig:disorder_realization} show the effects of finite disorder realizations for $m/g=1.1$, with different disorder realizations of 8 and 32 and the same initial states of 4. To exclude impacts from other aspects, we set the evolution time to infinity and consider two different lattice sizes, i.e. the size of the experimental system $N_c=L/2=16$ and a larger size of $N_c=55$. We can see that the error of the mean for both dynamical observables decreases as the number of disorder realizations increase, while the values of the mean are quantitatively consistent with each other. 

\subsection{Effect of finite system sizes}

\begin{figure}
	\centering
	\includegraphics[width=\linewidth]{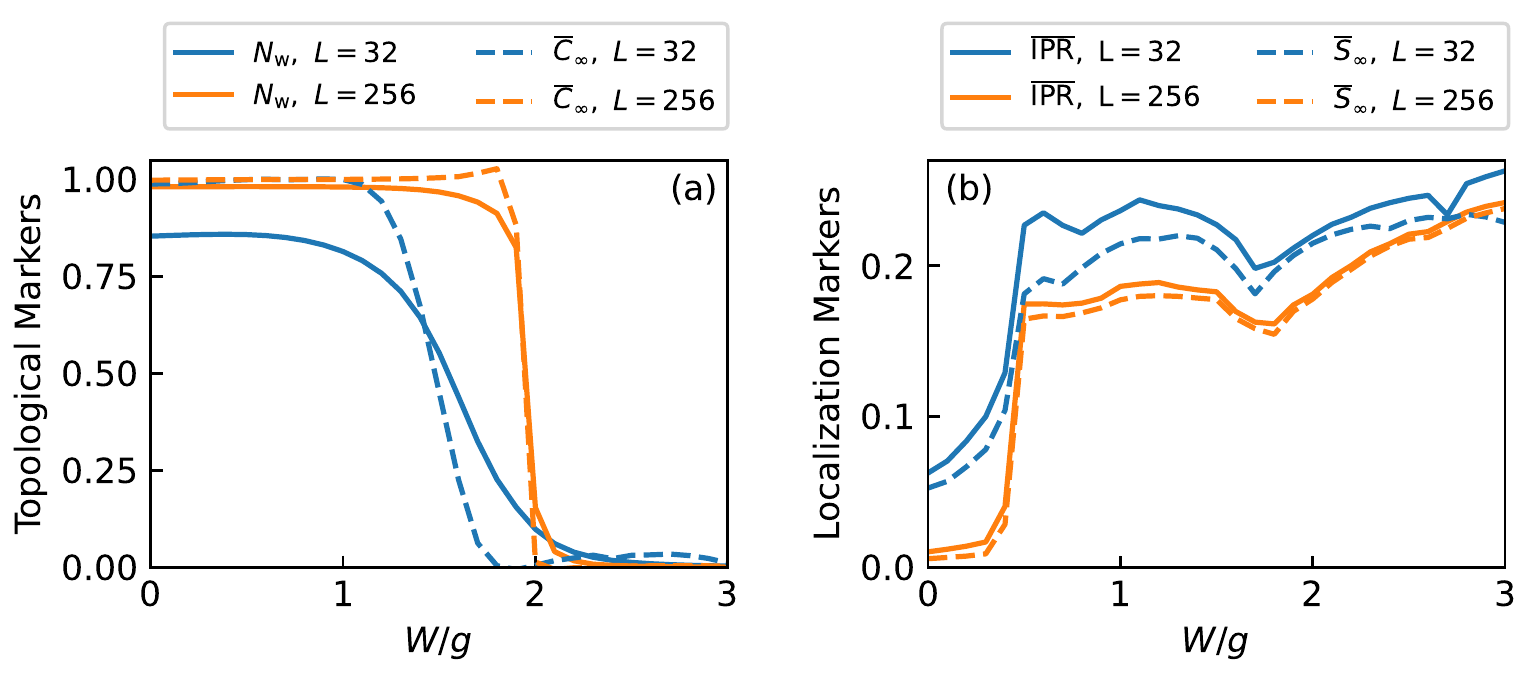}
	\caption{Finite-size effects. (a) Comparison between the winding number $N_{\rm w}$ and the mean chiral displacement $\overline{C}_\infty$. (b) Comparison between the spectrally averaged IPR $\overline{\rm IPR}$ and the mean survival probability $\overline{S}_\infty$. Here the intra-cell tunneling is fixed at $m/g=0.5$. The numerical results for $L=2N_c=32$ follow the experimental protocol, while the results for $L=256$ are obtained with 32 different initial states, i.e. single-excitation states with the excitation placed in the 16 unit cells in the middle of the chain.}
	\label{fig:finite_size_effects}
\end{figure}

\begin{figure}[b]
	\centering
	\includegraphics[width=\linewidth]{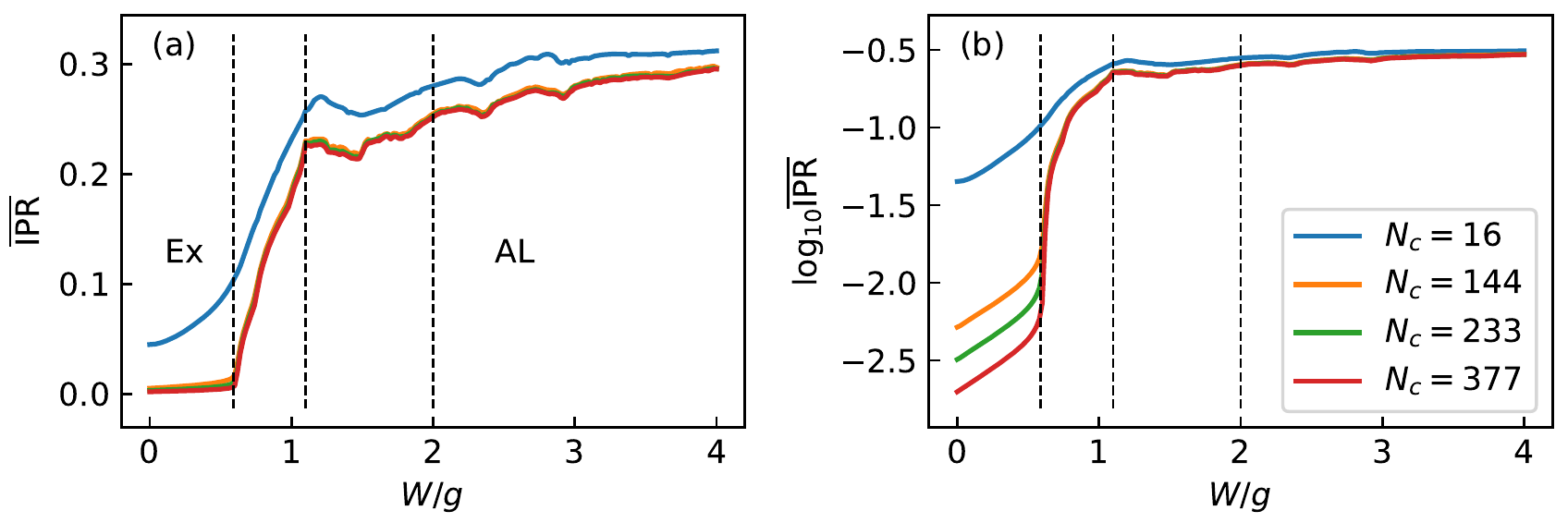}
	\caption{Calculated spectrally averaged IPR versus $W/g$ for the gSSH model with $m/g = 1.1$ and different system size $N_c$ plotted in (a) linear and (b) semi-log scales. Three vertical lines indicate phase boundaries 1, 2, and 3 in Fig. 4(d) of the main text. In the left and right regions defined by the three lines, mixed extended-localized states and mixed critical-localized states exist, respectively, while in the left-most and right-most regions are the extended (Ex) and Anderson localized (AL) states as indicated.}
	\label{fig:IPRs_extended_vs_localized}
\end{figure}

In Fig.~\ref{fig:finite_size_effects}, the numerial results show the comparison between the theoretical markers ($N_{\rm w}$, $\overline{\rm IPR}$) and the dynamical observables ($\overline{C}_{\infty}$, $\overline{S}_{\infty}$). We observe that the dynamical observables keep the same tendency as their corresponding theoretical results even for the small system size of $N_c=16$ and their discrepancies shrink as the system size increases. For relatively large system sizes, e.g. $N_c=128$, we can observe abrupt changes across phase boundaries. The numerical results for $N_c=16$ are obtained according to the same protocol as in the experiment, while those for $N_c=128$ are obtained with $8\times 32$ different quantum-walk realizations, with 8 different disorder phases and 32 different excitation positions in the middle of the chain.

In Fig.~\ref{fig:IPRs_extended_vs_localized}, we show numerical results of the spectrally averaged IPR as a function of the quasiperiodic disorder strength for the gSSH model with $m/g=1.1$ and different system sizes. In the extended phase, $\overline{\rm IPR}$ is close to zero and decreases as the size of the system increases. Note that the value of $\overline{\rm IPR}$ keeps increasing as the quasiperiodic disorder strength increases in the Anderson localized phase. However, it remains smaller than 1 even in the strong disorder regime, reflecting that the localized states are not strictly localized to a single site and always have a finite localization length.


\subsection{Effect of finite evolution times}

In the above discussion, we consider the dynamical evolution in an infinitely long time period. In the experiment, we have the maximum evolution time of $\tau=1.5~\mu{\rm s}$, corresponding to $g\tau\simeq 14$ where $g=2\pi\times1.5$ MHz is the inter-cell tunneling strength used in the experiment. 

\begin{figure}[t]
	\centering
	\includegraphics[width=\linewidth]{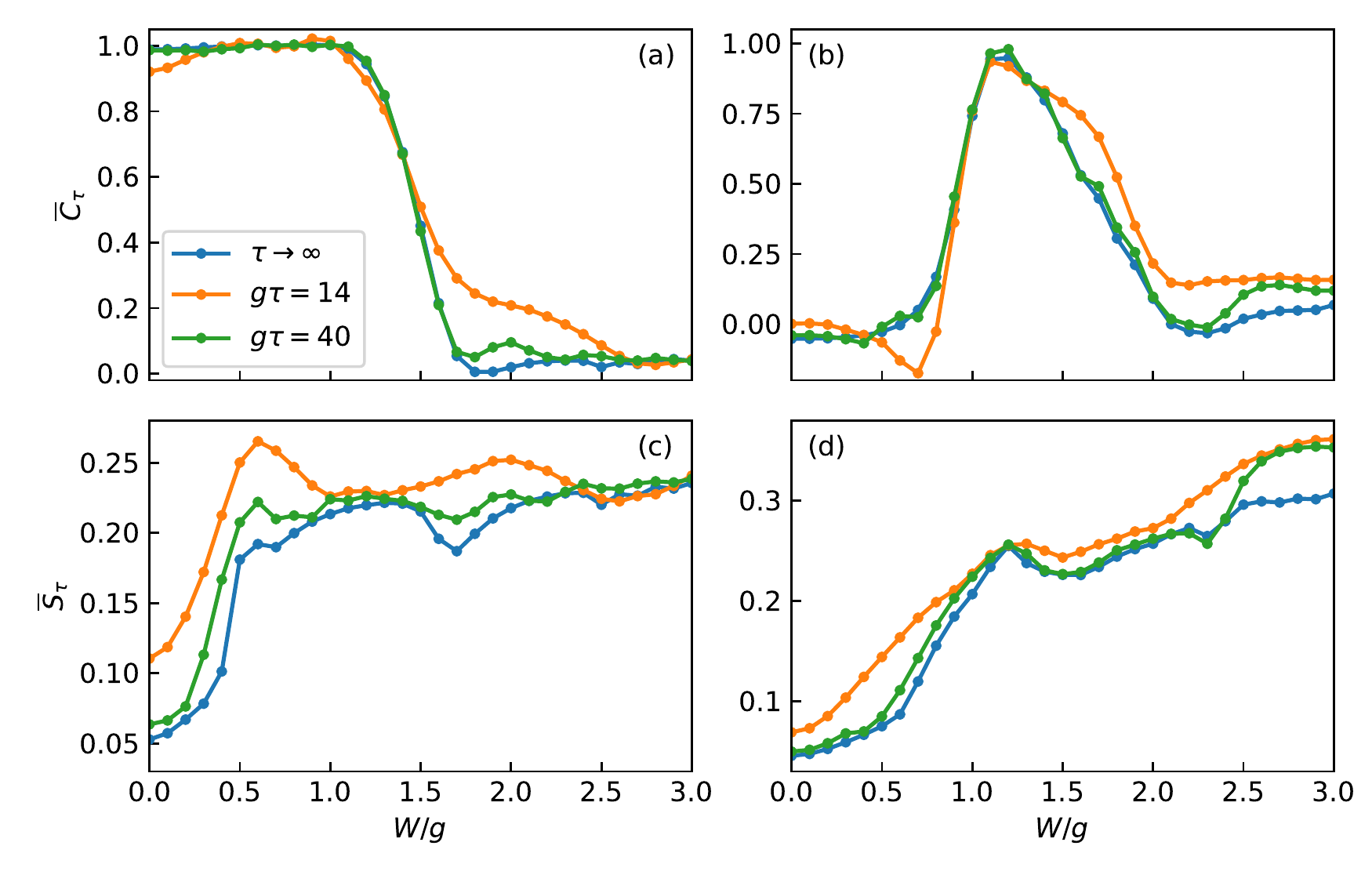}
	\caption{Finite-time effects. Mean chiral displacement measured at different evolution times for (a) $m/g=0.5$ and (b) $m/g=1.1$. The corresponding results for the mean survival probability are shown in (c) and (d). The number of unit cells is $N_c = 16$. For the finite evolution times, the numerical results are obtained by evolving the Schr\"{o}dinger equation following the experimental protocol, while the numerical results for the infinite-time cases are obtained by exact diagonalization.}
	\label{fig:finite_time_effects}
\end{figure}

In Fig.~\ref{fig:finite_time_effects}, we show the numerical results of the dynamical observables, i.e., $\overline{C}_\tau$ and $\overline{S}_\tau$, for systems with $N_c=16$ and different evolution times $g\tau=14$, $40$ and $\infty$. The finite-time results are qualitatively consistent with the infinite-time counterparts, indicating that they can be used experimentally for the study of the topological and localization properties.

\section{Phase diagram and phase boundaries}

The complete phase diagram of the gSSH model in Eqs.~(\ref{eq:hamil_gSSH}) and (\ref{eq:m_n}) studied in this work is shown in Fig.~4(d) in the main text. Although it is similar with the phase diagram presented in Ref.~\cite{tang2022topological}, it has essential differences that there exists a pure critical phase near the horizontal axis with $m \ll g$. Also, for $m < W$ there is a large parameter regime where the critical and Anderson localized states coexist, which leads to the existence of the generalized mobility edges.

In this section, we will first describe how to numerically determine the phase boundaries separating different bulk states. Next, we show the multifractality and similarity of the critical state seen from the energy spectrum and wave functions. Then we demonstrate that the topological phase boundaries can be obtained analytically, and numerically investigate the scaling behaviors and critical exponents across different phase boundaries. Finally, we discuss the effect of {\it inter}-cell quasiperiodic off-diagonal disorder to the observations of the critical states and TAI that are central in the present experiment. 

\subsection{Phase boundaries separating different bulk states}

The numerical procedure to determine the phase boundaries separating different bulk states has been extensively investigated in Ref.~\cite{tang2022topological}. With the appearance of the pure critical states and mixed critical-Anderson localized states, the phase boundary determination becomes more complicated, which is described below.

First of all, we note that the spectrally averaged IPR in Eq.~(\ref{eq:average_IPR}) is not enough to discriminate the pure extended or Anderson localized phase from their mixed phases. The solution is to follow the approach in Refs.~\cite{li2017mobility,li2020mobility,roy2021reentrant}, in which the spectral average of the normalized participation ratio (NPR) is first introduced:
\begin{eqnarray}
\overline{\rm NPR}\equiv L^{-1}\sum_n{\rm NPR}_n,~{\rm with}~{\rm NPR}_n=\left(L\times{\rm IPR}_n\right)^{-1},
\end{eqnarray}
and the phase boundary is determined by examining the quantity
\begin{eqnarray}
\eta \equiv \log_{10}\left(\overline{\rm IPR}\times\overline{\rm NPR}\right).
\end{eqnarray}

\begin{figure}[t]
	\centering
	\includegraphics[width=\linewidth]{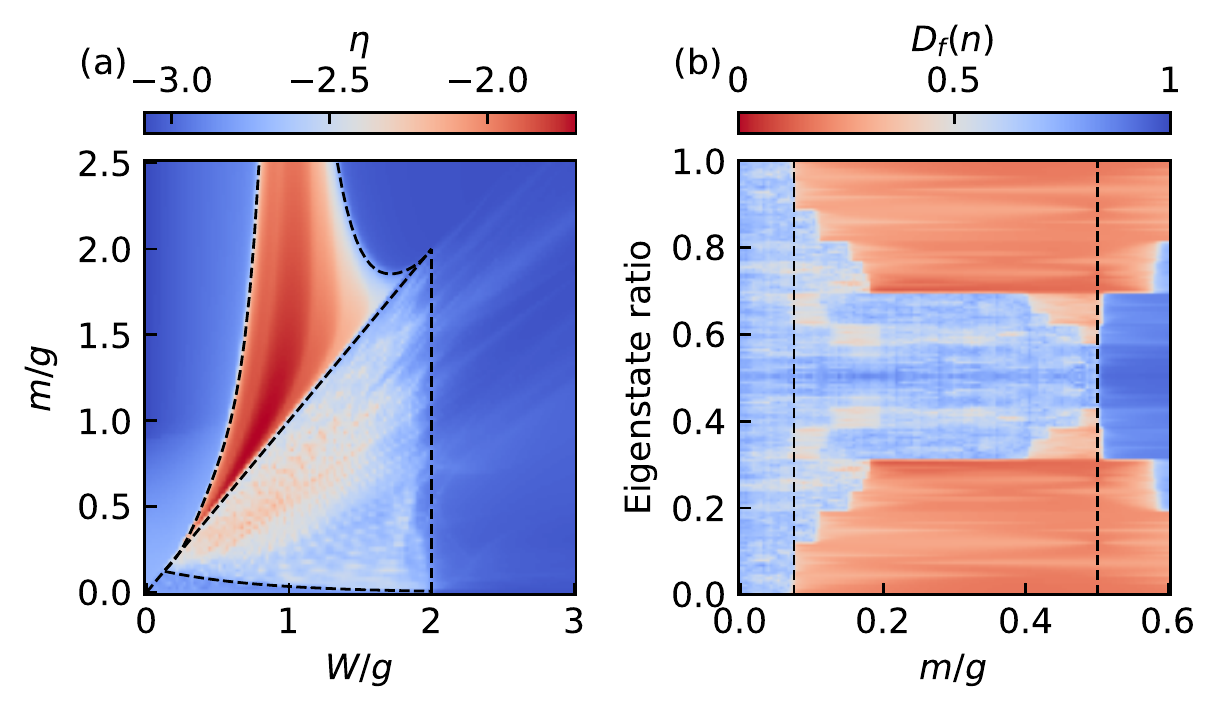}
	\caption{Localization phase boundaries. (a) Quantum $\eta$ as a function of the intra-cell tunneling $m$ and the quasiperiodic disorder strength $W$. (b) Eigenstate fractal dimension $D_f(n)\equiv -\log{\rm IPR}_n/\log L$ as a function of $m$ and the eigenstate ratio $n/L$ with $n=0,\ldots,L-1$ for the gSSH models with fixed $W/g=0.5$. In both panels, the results are obtained by diagonalizing the gSSH Hamiltonians with the chain length being $L=2N_c=1220$ and the black dashed lines are boundaries that separate regimes with different transport properties.}
	\label{fig:phase_boundaries}
\end{figure}

The numerical results of the quantity $\eta$ for the gSSH model with the chain length $L=1220$ are shown in Fig.~\ref{fig:phase_boundaries}(a). It is clear that the parameter plane can be divided into several regimes depending on the value of $\eta$, with $\eta\simeq0$ for the extended as well as the Anderson localized phases. There also exist the intermediate regimes lying between the extended and the AL phases. To explore the bulk state properties of these intermediate phases, we calculate the fractal dimension $D_f(n)\equiv -\log{\rm IPR}_n/\log L$ for all the eigenstates of the gSSH model, taking $W/g=0.5$ and $m/g\in[0,0.6]$ as an example and present the results in Fig.~\ref{fig:phase_boundaries}(b). The range of $m/g$ is chosen such that it goes across all the intermediate phases and the value of $D_f(n)$ can be used to discriminate the bulk state property of the $n$-th eigenstate $\ket{\phi_n}$, namely $D_f(n)\simeq 1$ or 0 for $\ket{\phi_n}$ being extended or Anderson localized, otherwise it is critically localized. It is clear that the spectrum of the three regimes from left to right are critical, mixed critical and Anderson localized, and mixed extended and Anderson localized, respectively. In these cases, there exist not only the conventional mobility edges, but also generalized mobility edges which separate critically localized and Anderson localized eigenstates in certain parameter regimes, as shown in Fig.~4(e) of the main text.

\subsection{Critical phase: multifractality and self-similarity}

\begin{figure*}[t]
	\centering
	\includegraphics[width=0.9\textwidth]{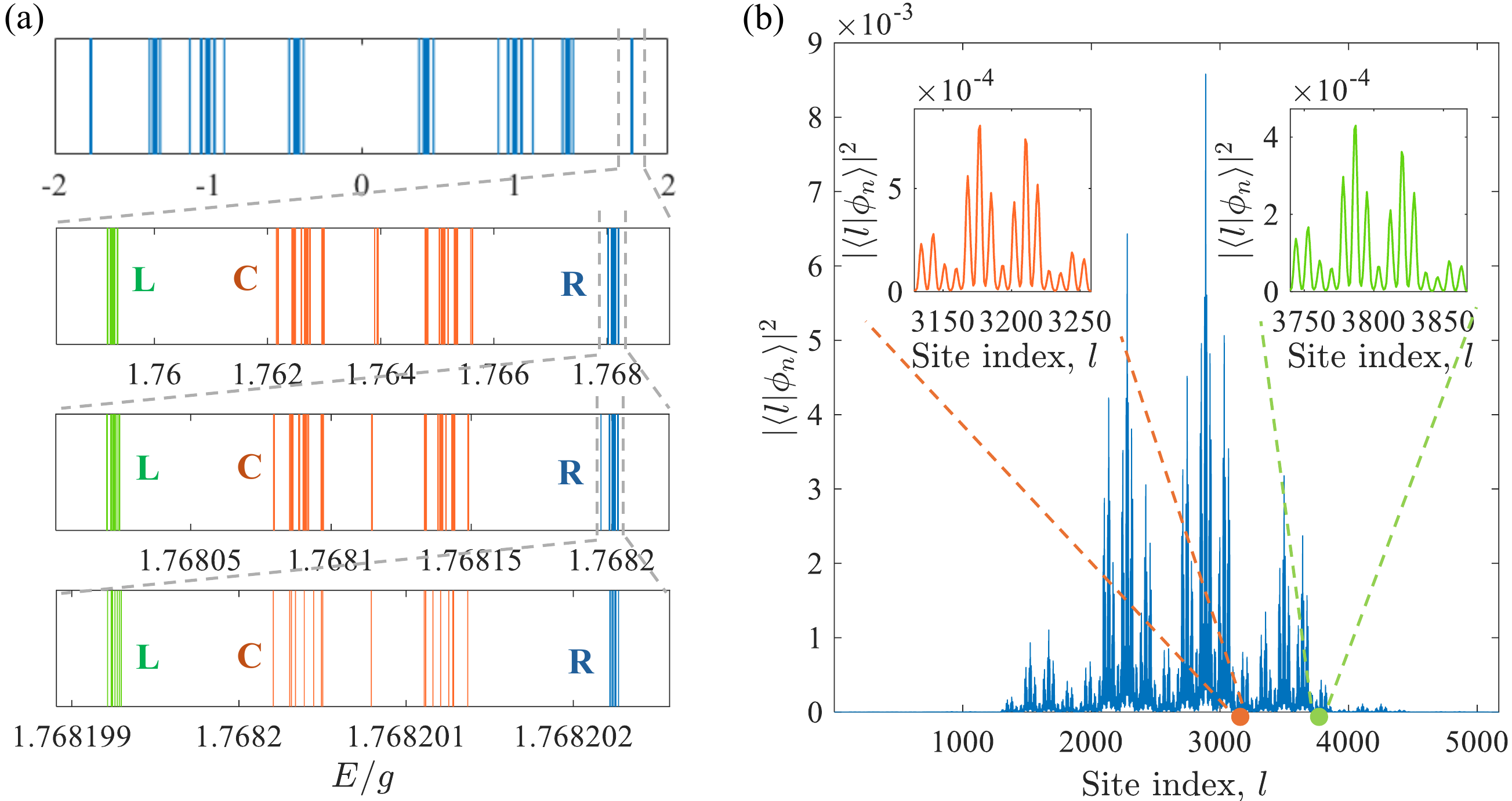}
	\caption{Self-similarities of (a) energy spectrum and (b) wave function when the system is in the fully critical phase under periodic boundary condition. The parameters are $L=2N_c=5168$, $m/g=0$, $W/g=1$, and $\beta=(\sqrt{5}-1)/2$.
	}\label{similarity}
\vspace{8mm}
	\includegraphics[width=0.9\textwidth]{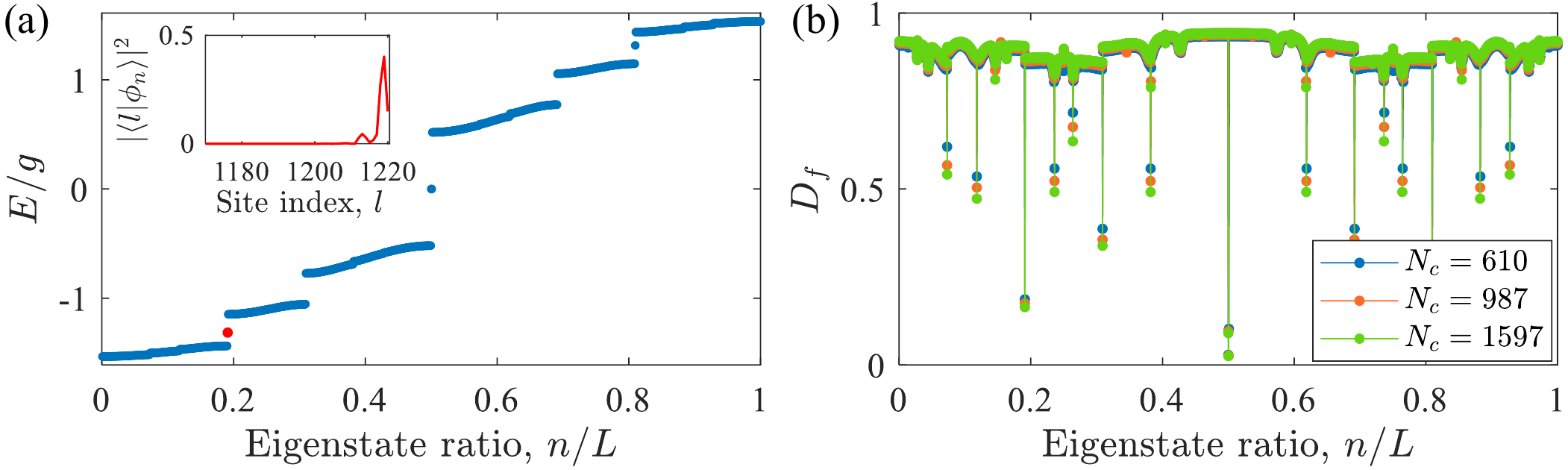}
	\caption{(a) Eigenenergy spectrum under open boundary condition for $N_c=610$. The inset shows the density distribution of the $233$th eigenstate marked by red point in the spectrum. (b) $D_f$ of the spectrum for different system sizes $N_c$ = 610, 987, and1597. Other parameters are $m/g=0.5$ and $W/g=0.3$.}
	\label{ingapstate}
\end{figure*}

The spectra of the critical states exhibit self-similarity in our quasiperiodic gSSH model under both the periodic and open boundary conditions. Figure~\ref{similarity}(a) shows the energy spectrum of $\hat{H}_{\rm gSSH}$ for $m/g=0$ and $W/g=1$ under periodic boundary condition, for which all the eigenstates are critically localized. Considering the right-most top band, one finds three clusters of subbands (denoted by L, C ,and R), which can be seen clearly by expanding the spectrum. Each of the three clusters further splits into three clusters. Such splitting is continued ad infinitum in the thermodynamic limit for irrational modulation parameter $\beta$, which indicates the self-similar structure of the spectrum~\cite{Ketzmerick1998}. In addition, the fractal wave functions of critical states also exhibit self-similarity \cite{Shimasaki2024}. For instance, the density distribution of the lowest-energy wave function with self-similar structure can be seen in Fig.~\ref{similarity} (b).

Under open boundary condition, we find some states localized near boundaries of the irrational lattice, which are lying between subbands of the fractal spectrum. Figure~\ref{ingapstate}(a) shows an example. These in-gap localized states are not protected by the chiral symmetry, thus different from the topological zero-energy edge states. In Fig.~\ref{ingapstate} (b), we show $D_f$ for different system sizes $N_c$ = 610, 987, and 1597, which indicates that these localized states exist in thermodynamic limit.

\begin{figure*}[t]
	\centering
	\includegraphics[width=0.99\textwidth]{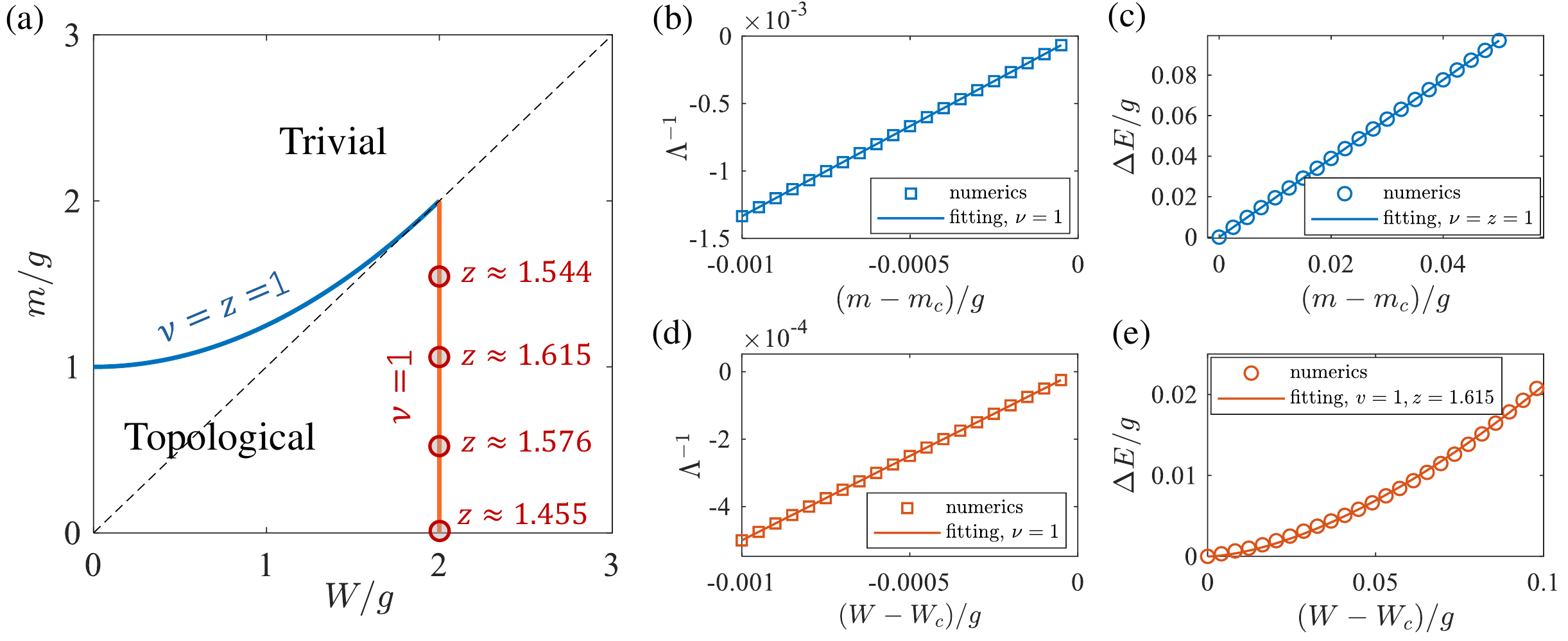}
	\caption{(a) Blue and red lines denote exact topological boundaries given by Eq. (\ref{topoboundary}), which are separated by the black dashed line of $m=W$. The critical exponents are numerically obtained as $\nu=z=1$ for the blue boundary line, while $\nu=1$ and varying $z$ denoted for the red boundary line. Numerical scaling of $\Lambda^{-1}$ (b,d) and $\Delta E$ (c,e) with fitting exponents. Other parameters are $W/g=1$ in (b,c) and $m/g=1$ in (d,e) with $L=2N_c=1220$.
	}\label{phaseboundary}
\end{figure*}

\subsection{Topological phase boundaries and scaling behaviors}

In this subsection, we show that the topological phase boundaries in Fig.~4(d) of the main text can be derived analytically. Due to the gap-closing property at the topological transition point, the localization length $\Lambda$ of zero-energy modes under open boundary condition diverges \cite{MondragonShem2014}, which corresponds to $\Lambda^{-1}\to0$. The zero-energy eigenstate $\phi_{\rm edge}=\{\phi_{1,A}, \phi_{1,B}, \cdots \phi_{N,A}, \phi_{N,B}\}^T$ can be obtained by solving $\hat{H}_{\rm gSSH}\ket{\phi_{\rm edge}} =0$ as
\begin{equation}
	\phi_{n,A}=(-1)^{n}\prod_{l=1}^{n}\frac{m_{l}}{g}\phi_{1,A}.
\end{equation}
Hence, $\Lambda^{-1}$ for zero-energy modes reads
\begin{equation}\label{Lambda1}
\begin{aligned}
	\Lambda^{-1}
	&=\lim\limits_{N_c\to\infty}\frac{1}{N_c}\ln|\frac{\phi_{N_c,A}}{\phi_{1,A}}|\\
	&=\lim_{N_c\to\infty}\frac{1}{N_c}\sum_{n=1}^{N_c}\left(\ln| m_n| - \ln|g| \right).
\end{aligned}
\end{equation}
According to Birkhoff's ergodic theorem~\cite{MondragonShem2014} and denoting $x\doteq2\pi\beta n+\delta$, we can rewrite Eq. (\ref{Lambda1}) as
\begin{equation}\label{Lambda2}
	\Lambda^{-1}=\frac{1}{2\pi}\int_{-\pi}^\pi dx  \ln \lvert { m+W\mathrm{cos}x} \rvert-\ln|g|.
\end{equation}
It can be solved as
\begin{equation}
	\Lambda^{-1}=\begin{cases} \ln\frac{m+\sqrt{m^2-W^2}}{2g}, &m>W;\\
		\ln\frac{W}{2g}, &m<W.  \end{cases}
\end{equation}
The topological transition points $m_c$ and $W_c$ for $m>W$ and $m<W$ can be obtained by letting $\Lambda^{-1}=0$, which read:
\begin{equation}\label{topoboundary}
\begin{aligned}
	&m_c=g+W^2/4g,  ~~~~~    &m>W;\\
	&W_c=2g,   &m<W.
\end{aligned}
\end{equation}
The analytical results are plotted in Fig. \ref{phaseboundary} (a), which are perfectly consistent with the topological phase boundaries determined numerically.

Our analytical result indicate different topological behaviors for $m>W$ and $m<W$, which correspond to two parameter regions separated by the line $m=W$ in the $m$-$W$ plane, as shown in Fig. \ref{phaseboundary} (a). To further reveal the difference of topological transitions for $m>W$ and $m<W$, we numerically study the correlation length and dynamic critical exponents. The correlation length (of the ground state at half-filling of free fermions) near topological transition point can be estimated by the localization length of zero-energy modes~\cite{MondragonShem2014,Chandran2017}. Near the critical points $m_c$ and $W_c$ and in the large $L$ limit, the inverse of the localization length $\Lambda^{-1}$ approaches zero with the scaling forms:
\begin{equation}
	\begin{aligned}
		&\Lambda^{-1}\sim |m-m_c|^{\nu},  ~~~~~    &m>W; \\
		&\Lambda^{-1}\sim |W-W_c|^{\nu},   &m<W.
	\end{aligned}
\end{equation}
Here $|m-m_c|$ and $|W-W_c|$ denote the deviation from phase boundary, and $v$ denotes the correlation-length critical exponent. The dynamic critical exponent $z$ can be further obtained from the scaling forms of the energy gap $\Delta E$ at half filling, which reads
\begin{equation}
	\begin{aligned}
	&\Delta E\sim |m-m_c|^{{\nu}z},  ~~~~~    &m>W; \\
	&\Delta E\sim |W-W_c|^{{\nu}z},   &m<W.
\end{aligned}
\end{equation}

By using the scaling functions, we can numerically extract the critical exponents $\nu$ and $z$. We obtain $\nu=z=1$ for topological critical line of $m_c=g+W^2/4g$ for $m>W$. This indicates that the topological transition in the $m>W$ region belongs to the same university class at disorder-free case with $m_c=1$ and $W=0$. For the critical line of $W_c=2$ in the $m<W$ region, we find $\nu=1$ and non-universal values of $z$, such as $z\approx\{1.455,1.576,1.615,1.544\}$ for $m/g=\{0,0.5,1.0,1.5\}$, respectively, which indicate that the topological transition in this relatively strong disorder regime is not equivalent to that disorder-free case. Figures~\ref{phaseboundary} (b) and (c), and (d) and (e) show two examples of the scaling of $\Lambda^{-1}$ and $\Delta E$. In a word, our analytical and numerical results demonstrate that topological transitions above and below the diagonal line $m=W$ in the $m$-$W$ plane belong to two different classes. Notably, in the region above and below the diagonal line $m=W$, the critical states are absent and present, respectively, thus the non-universal dynamic exponents for the topological transition at $W_c=2$ may result from the transition between the critical and localized states in the middle of the energy spectrum.

\begin{figure*}[t]
	\centering
	\includegraphics[width=0.99\textwidth]{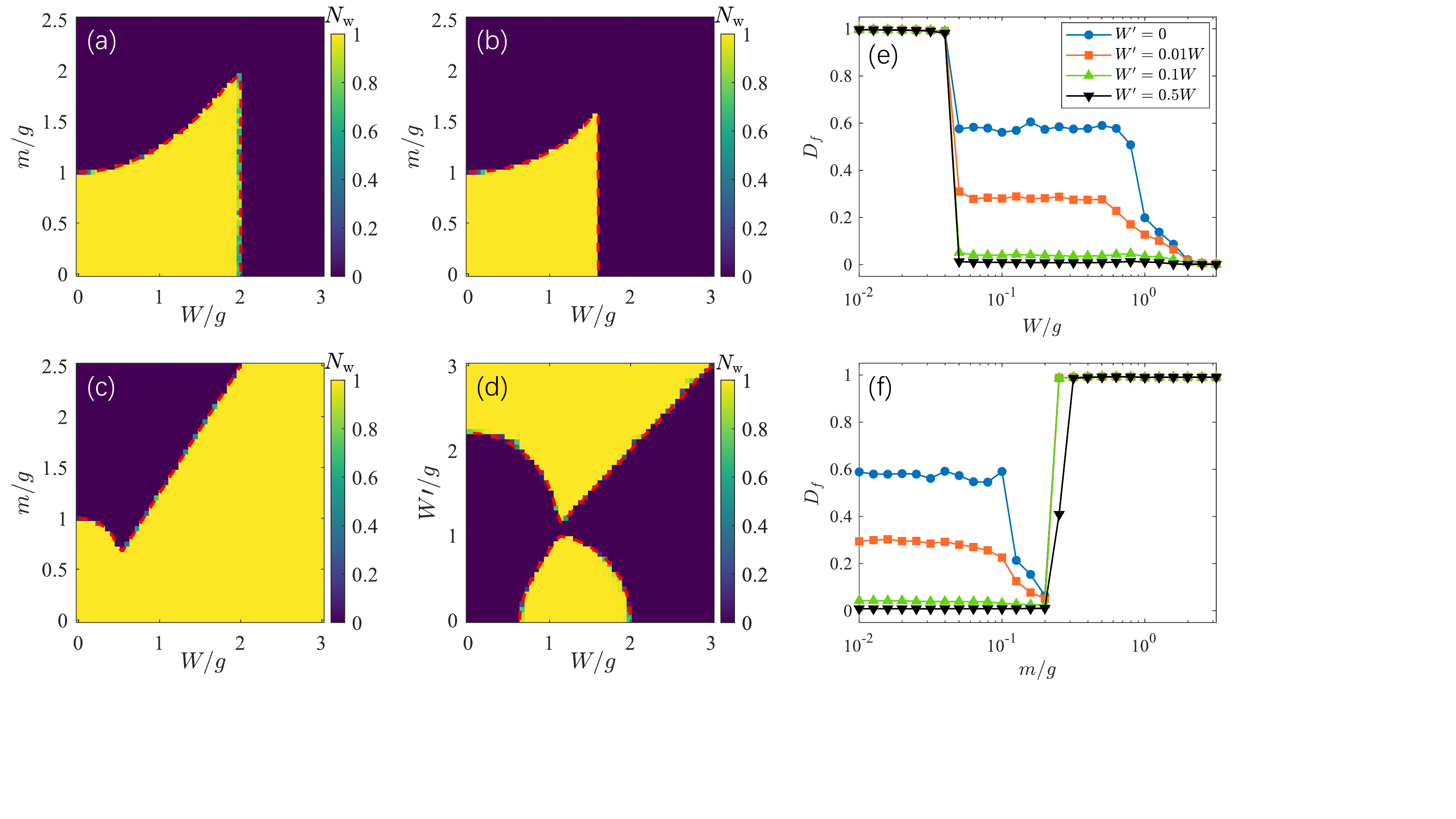}
	\caption{Winding number $N_{\rm w}$ as functions of $W/g$ and $m/g$ for (a) $W'=0$, (b) $W'=0.5W$ and (c) $W'=2W$. (d) $N_{\rm w}$ as a function of $W/g$ and $W'/g$ for $m/g=1.1$. The red dash lines in (a-d) denote where the localization length of zero-energy modes diverges for the system size $N_c=610$. $D_f$ as a function of (e) $W/g$ for fixed $m/g=0.05$, and (f) $m/g$ for fixed $W/g=0.2$. Here $D_f$ is obtained by fitting the results of $\overline{\rm IPR}$ with system sizes $N_c$ = 610, 987, and 1597.
	}\label{intercell-disorder}
\end{figure*}

We note that the phase boundaries between topologically nontrivial and trivial phases are usually obtained by numerically evaluating the real-space representation of the winding number in Eq.~(\ref{eq:winding_number}) for the gSSH model with a large enough chain length~\cite{tang2022topological}. The winding number $N_{\rm w}$ will change abruptly from 1 to 0 across the phase boundaries, as can be seen in Fig.~\ref{intercell-disorder}(a) calculated with $L=2N_c=1220$. 

%

\subsection{Effect of inter-cell quasiperiodic disorders}

We now discuss the effect of inter-cell quasiperiodic disorder to the observations of TAIs and critical states. The total Hamiltonian reads
\begin{equation}
	\hat{H}/\hbar= \hat{H}_{\rm gSSH}/\hbar+ \sum_{n=1}^{N_{c}-1}       W^\prime \cos (2n \pi \beta + \delta) \hat{a}_{n+1}^{\dagger}\hat{b}_{n},
\end{equation}
with $\hat H_{\rm gSSH}$ given by Eq.~(\ref{eq:hamil_gSSH}), where $W^\prime$ denotes the quasiperiodic disorder strength of the inter-cell hopping. Figures~\ref{intercell-disorder}(a)-(c) show the winding number $\nu$ as functions of $W/g$ and $m/g$ for various $W^\prime$. One can find that the additional inter-cell disorder can significantly affect the topological phase diagram. When $W^\prime/W<1$ such as $W^\prime=0.5W$ in Fig. \ref{intercell-disorder}(b), the topological region shrinks compared with the case of $W^\prime/g=0$ in Fig. \ref{intercell-disorder}(a). When $W^\prime/W>1$ such as $W^\prime=2W$ in Fig. \ref{intercell-disorder}(c), the disorder-induced topological region enlarges with increasing disorder strength. This can be understood by the fact that the disorder-renormalized inter-cell hopping strength is always larger than the intra-cell hopping in the large $W^\prime=2W$ region, where the topological phase preserves. In Fig. \ref{intercell-disorder}(d), we show the topological phase diagram in the $W$-$W^\prime$ plane with fixed $m/g=1.1$. One can find the TAI region shrinks and disappears when increasing $W^\prime$ up to $W^\prime/g\approx1$, and then the system reenters the topological phase. We further numerically determine the topological phase boundaries by the divergence of the localization length of zero-energy modes, shown in Figs. \ref{intercell-disorder}(a)-(d) as dashed lines, which are consistent with those seen from the winding number. However, in the presence of additional inter-cell disorder, it is difficult to analytically obtain the exact topological phase boundaries. We also consider the effect of inter-cell disorders on the localization. As excepted, this additional disorder term tends to localize the extended and critical states.
Figures \ref{intercell-disorder}(e) and (f) show the average fractal dimension $D_f$ as functions of $W/g$ for fixed $m/g=0.05$ and $m/g$ for fixed $W/g=0.2$, respectively. The results indicate that the extended phase with $D_f\approx1$ and critical phase with $D_f\approx0.6$ will become localized phase when increasing the inter-cell disorder strength.

\section{Relations with other models}

In this section, we discuss some interesting relations of the gSSH model considered in this work, given by Eqs.~(\ref{eq:hamil_gSSH}) and (\ref{eq:m_n}), with the generalized AAH (gAAH) and Ising models, either in limiting cases or under exact mapping, which help to gain more insights into the rich structure of the topology-localization phase diagram. 

\subsection{Limit $m\gg W, g$}

First, we consider the limiting case of $m\gg W,~g$. Under this condition, the fermionic modes inside a unit cell would hybridize to diagonalize the intra-cell tunneling, and the inter-cell tunneling terms would play the role of perturbation.

We define a set of annihilation operators $\hat c_n$ and $\hat d_n$, which diagonalize the intra-cell tunneling,
\begin{eqnarray}
\hat c_n=\frac{1}{\sqrt{2}}\left(\hat a_n-\hat b_n\right),\quad \hat d_n=\frac{1}{\sqrt{2}}\left(\hat a_n+\hat b_n\right),
\end{eqnarray}
and rewrite the Hamiltonian in terms of these operators,
\begin{eqnarray}
\hat H_{\rm gSSH}\!\!&=&\!\!\sum_n \left[m+W\cos\left(2\pi\beta n+\delta\right)\right]\left(\hat d_n^\dag\hat d_n-\hat c_n^\dag\hat c_n\right)\\
&&\!\!+\frac{g}{2}\sum_n\left(\hat d_n^\dag\hat d_{n+1}+\hat d^\dag_{n+1}\hat d_n\right)\nonumber\\
&&\!\!-\frac{g}{2}\sum_n\left(\hat c_n^\dag\hat c_{n+1}+\hat c^\dag_{n+1}\hat c_n\right)\nonumber\\
&&\!\!-\frac{g}{2}\sum_n\left(\hat c_n^\dag\hat d_{n+1}+\hat d_{n+1}^\dag\hat c_n-\hat d_n^\dag\hat c_{n+1}-\hat c_{n+1}^\dag\hat d_n \right).\nonumber
\end{eqnarray}
This is equivalent to two copies of the AAH model, of which the average values of the on-site potential are $m$ and $-m$, and the coupling strength between these two AAH model is characterized by $g/2$. In the limit of $m\gg W, g$, the couplings between the two AAH lattices are much smaller than the energy splitting, i.e. $\frac{g}{2}\ll 2m$, such that the inter-lattice coupling terms can be safely neglected. Hence the model is mapped to two decoupled AAH models, i.e.
\begin{eqnarray}
& &\lim_{m\rightarrow\infty}\hat H_{\rm gSSH}(m, W, t)\equiv \hat H_{\rm AAH}\\
&=&-W\sum_n\cos\left(2\pi\beta n+\delta\right)\hat c_n^\dag\hat c_n\nonumber\\
&&-\frac{g}{2}\sum_n\left(\hat c_n^\dag\hat c_{n+1}+\hat c_{n+1}^\dag\hat c_n\right),\nonumber
\end{eqnarray}
with the extended-to-localized phase transition happens at the critical value of $W/g=1$. 

\subsection{Limit $g \gg W, m$}

Under the condition $g \gg W, m$, it is reasonable to introduce operators that diagonalize the inter-cell tunneling, i.e.
\begin{eqnarray}
\hat c_n=\frac{1}{\sqrt{2}}\left(\hat b_n-\hat a_{n+1}\right),\quad \hat d_n=\frac{1}{\sqrt{2}}\left(\hat b_n+\hat a_{n+1}\right).
\end{eqnarray}
With the inverse relations
\begin{eqnarray}
\hat b_n=\frac{1}{\sqrt{2}}\left(\hat d_n+\hat c_n\right),\quad \hat a_{n+1}=\frac{1}{\sqrt{2}}\left(\hat d_n-\hat c_n\right),
\end{eqnarray}
the Hamiltonian in Eq.~(\ref{eq:hamil_gSSH}) can be rewritten as 
\begin{eqnarray}
\hat H_{\rm gSSH}&=&-g\sum_n \hat c_n^\dag\hat c_n-\sum_n\frac{m_n}{2}\left(\hat c_{n-1}^\dag\hat c_n+\hat c_n^\dag\hat c_{n-1}\right)\\
&&+g\sum_n \hat d_n^\dag\hat d_n+\sum_n\frac{m_n}{2}\left(\hat d_{n-1}^\dag\hat d_n+\hat d_n^\dag\hat d_{n-1}\right)\nonumber\\
&&+\sum_n\frac{m_n}{2}\left(\hat d_{n-1}^\dag\hat c_n-\hat c_{n-1}^\dag\hat d_n+{\rm h.c.}\right).\nonumber
\end{eqnarray}
This is equivalent to two weakly coupled bands of the off-diagonal AAH model. In the limit of $g\gg m, W$, the coupling term in the last line can be safely neglected. As long as this condition is valid, the phase diagram is the same as that of the off-diagonal AAH model~\cite{chang1997multifractal}, with the critical point of $m=W$ for the transition from the extended phase to the critical phase.

\subsection{Quasiperiodic transverse-field Ising model}

To transform the gSSH Hamiltonian in Eq.~(\ref{eq:hamil_gSSH}) into the Majorana representation, we first introduce the Majorana fermion operators,
\begin{eqnarray}
&\hat\gamma_{2n}=\hat a_n^\dag+\hat a_n,\quad &\hat\gamma_{2n+1}=i\left(\hat a_n^\dag-\hat a_n\right),\\
&\hat\tau_{2n}=\hat b_n^\dag+\hat b_n,\quad &\hat\tau_{2n+1}=i\left(\hat b_n^\dag-\hat b_n\right),\nonumber
\end{eqnarray}
with the inverse relation
\begin{eqnarray}
&\hat a^\dag=\frac{1}{2}\left(\hat\gamma_{2n}-i\hat\gamma_{2n+1}\right),~ &\hat a_n=\frac{1}{2}\left(\hat\gamma_{2n}+i\hat\gamma_{2n+1}\right),\\
&\hat b^\dag=\frac{1}{2}\left(\hat\tau_{2n}-i\hat\tau_{2n+1}\right),~ &\hat b_n=\frac{1}{2}\left(\hat\tau_{2n}+i\hat\tau_{2n+1}\right).\nonumber
\end{eqnarray}
Note that the commutation relations of the Majorana operators have the forms
\begin{eqnarray}
\left\{\hat\gamma_i,\hat\gamma_j\right\}=2\delta_{ij},\quad \left\{\hat\tau_i,\hat\tau_j\right\}=2\delta_{ij},\quad\left\{\hat\gamma_i,\hat\tau_j\right\}=0.
\end{eqnarray}
The gSSH Hamiltonian in Eq.~(\ref{eq:hamil_gSSH}) can be rewritten in terms of the Majorana operators,
\begin{eqnarray}
\hat H_{\rm gSSH}&=&\sum_n\frac{im_n}{2}\left(\hat\gamma_{2n}\hat\tau_{2n+1}-\hat\gamma_{2n+1}\hat\tau_{2n}\right)\\
&&+\frac{ig}{2}\sum_n\left(\hat\tau_{2n}\hat\gamma_{2n+3}-\hat\tau_{2n+1}\hat\gamma_{2n+2}\right)\nonumber\\
&=&\sum_n\left(\frac{im_n}{2}\hat\gamma_{2n}\hat\tau_{2n+1}-\frac{ig}{2}\hat\tau_{2n+1}\hat\gamma_{2n+2}\right)\nonumber\\
&&-\sum_n\left(\frac{im_n}{2}\hat\gamma_{2n+1}\hat\tau_{2n}-\frac{ig}{2}\hat\tau_{2n}\hat\gamma_{2n+3}\right).\nonumber
\end{eqnarray}
Moreover, the Majorana fermionic system can be mapped to a spin system with the Jordan-Wigner transformation. Specifically, we define the following mapping:
\begin{eqnarray}
&\hat\gamma_{2n} = \prod_{j<n}\hat\sigma_j^z\hat\sigma_n^x, \quad &\hat\tau_{2n+1} = \prod_{j<n}\hat\sigma_j^z\hat\sigma_n^y,\\
&\hat\gamma_{2n+1} = \prod_{j<N+n}\hat\sigma_j^z\hat\sigma_{N+n}^x,\quad &\hat\tau_{2n} = \prod_{j<N+n}\hat\sigma_j^z\hat\sigma_{N+n}^y.\nonumber
\end{eqnarray}
With these, the Hamiltonian can be rewritten in terms of the spin operators as:
\begin{eqnarray}
\hat H_{\rm gSSH}&=&-\sum_n\left(\frac{m_n}{2}\hat\sigma_n^z-\frac{g}{2}\hat\sigma_n^x\hat\sigma_{n+1}^x\right)\\
&&+\sum_n\left(\frac{m_n}{2}\hat\sigma_{N+n}^z-\frac{g}{2}\hat\sigma_{N+n}^x\hat\sigma_{N+n+1}^x\right),\nonumber
\end{eqnarray}
which corresponds to two mutually decoupled transverse-field Ising chains with quasi-periodically modulated transverse field.

\subsection{2D tight-binding model}

\begin{figure*}
    \centering
    \includegraphics[width=0.9\linewidth]{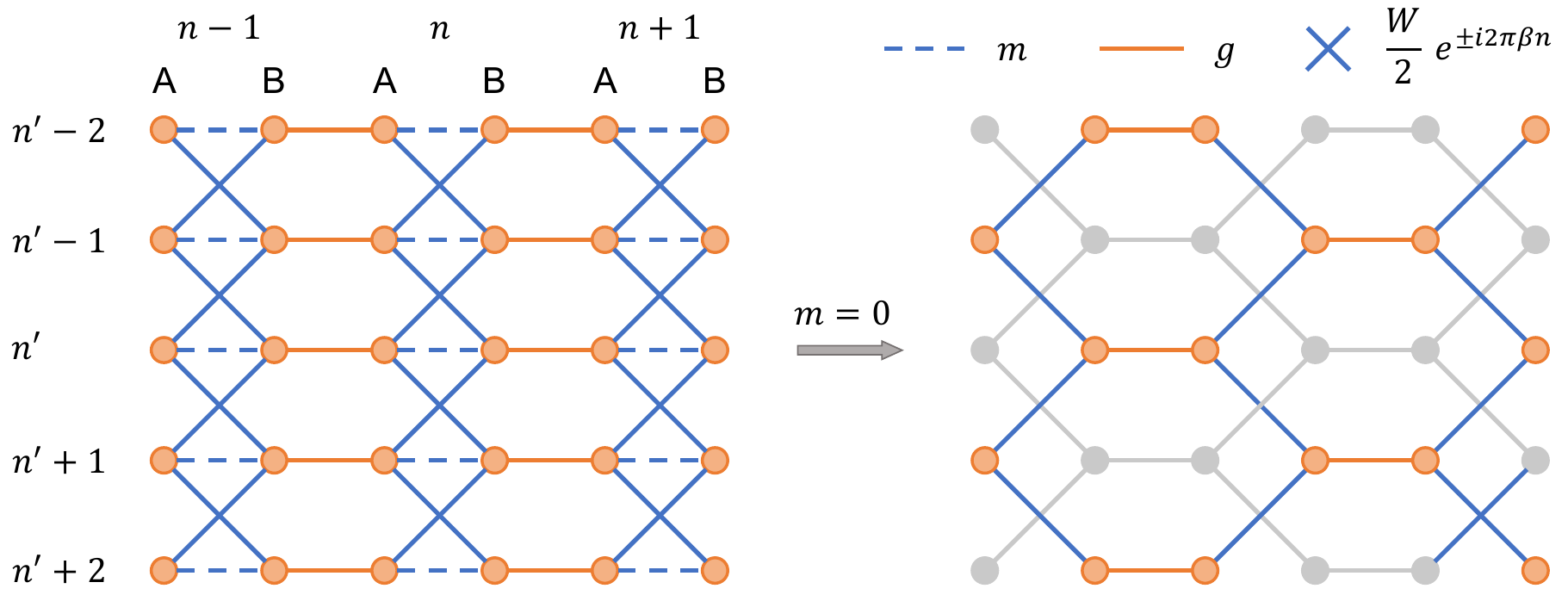}
    \caption{Left: Schematic plot of the 2D tight-binding model in Eq.~(\ref{eq:H_2D}). Right: When $m=0$, the 2D square lattice reduces to two identical and mutually penetrating honeycomb lattices.}
    \label{fig:2D_model}
\end{figure*}

Consider a tight-binding model on a 2D square lattice with the lengths along $x$ and $y$ directions denoted $L_x=2N_c$ and $L_y$. We will show that the following Hamiltonian
\begin{eqnarray}
\hat H_{\rm 2D}&=&m\sum_{n,n'}\left(\hat a^\dag_{n,n'}\hat b_{n,n'}+\hat b^\dag_{n,n'}\hat a_{n,n'}\right)\label{eq:H_2D}\\
&&+g\sum_{n,n'}\left(\hat b^\dag_{n,n'}\hat a_{n+1,n'}+\hat a^\dag_{n+1,n'}\hat b_{n,n'}\right)\nonumber\\
&&+\frac{W}{2}\sum_{n,n'}\left(e^{-i2\pi\beta n}\hat a_{n,n'}^\dag\hat b_{n,n'-1}+{\rm h.c.}\right)\nonumber\\
&&+\frac{W}{2}\sum_{n,n'}\left(e^{i2\pi\beta n}\hat a_{n,n'}^\dag\hat b_{n,n'+1}+{\rm h.c.}\right) \nonumber
\end{eqnarray}
can be mapped to the gSSH model considered in this work. Here $m$ and $g$ are the intra- and inter-cell tunneling strength of the gSSH model along each row of the square lattice. There are also tunnelings between the next-nearest-neighbors among adjacent rows inside the same unit cell, with the strength denoted as $W/2$. The tight-binding model is depicted in the left panel of Fig.~\ref{fig:2D_model}.

The relation of the 2D model and the gSSH model in Eq.~(\ref{eq:hamil_gSSH}) can be seen from Fourier transform along the $y$ direction. For this, we define the Fourier transformed operators as
\begin{eqnarray}
\tilde a_{n,k'}&=&\frac{1}{\sqrt{L_y}}\sum_{n'}e^{-ik'n'}\hat a_{n,n'},\\ \tilde b_{n,k'}&=&\frac{1}{\sqrt{L_y}}\sum_{n'}e^{-ik'n'}\hat b_{n,n'},\nonumber
\end{eqnarray}
with the inverse transform of
\begin{eqnarray}
\hat a_{n,n'}&=&\frac{1}{\sqrt{L_y}}\sum_{k'}e^{ik'n'}\hat a_{n,k'},\\ 
\hat b_{n,n'}&=&\frac{1}{\sqrt{L_y}}\sum_{k'}e^{ik'n'}\hat b_{n,k'}.\nonumber
\end{eqnarray}
Then the 2D Hamiltonian can be decoupled as $\hat H_{\rm 2D}=\sum_{k'}\hat H_{k'}$, with each $\hat H_{k'}$ in the form
\begin{eqnarray}
\hat H_{k'}&=&m\sum_{n}\left(\tilde a^\dag_{n,k'}\tilde b_{n,k'}+\tilde b^\dag_{n,k'}\tilde a_{n,k'}\right)\\
&&+g\sum_{n}\left(\tilde b^\dag_{n,k'}\tilde a_{n+1,k'}+\tilde a^\dag_{n+1,k'}\tilde b_{n,k'}\right)\nonumber\\
&&+W\sum_{n}\cos\left(2\pi\beta n+k'\right)\left(\tilde a^\dag_{n,k'}\tilde b_{n,k'}+\tilde b^\dag_{n,k'}\tilde a_{n,k'}\right).\nonumber
\end{eqnarray}
It is easy to see that each $\hat H_{k'}$ corresponds to the gSSH model in Eq.~(\ref{eq:hamil_gSSH}), with the quasi-momentum $k'$ playing the role of the phase $\delta$ for generating different disorder realizations. Therefore, the 2D model in Eq.~(\ref{eq:H_2D}) and the gSSH model in Eq.~(\ref{eq:hamil_gSSH}) are analytically equivalent.

The equivalence between the two models can be used to explain the  critical-to-localized phase transition with $m$ = 0 in the phase diagram shown in Fig. 4(d) of the main text. As shown in the right panel of Fig.~{\ref{fig:2D_model}}, removing all the intra-cell tunnelings in the SSH models along all of the rows results in two sets of fully decoupled honeycomb lattices, with the magnitudes of the tunneling strengths of the three bounds leading to the same sites being $\frac{W}{2}$, $\frac{W}{2}$ and $g$, respectively. As discussed in Ref.~\cite{Chandran2017}, this structure leads to a critical-to-localized phase transition with the critical value of $W/g=2$, which is consistent with our numerical results of the gSSH model.

\bibliography{supp}